\documentclass[11pt]{article} 
\usepackage{epsfig,a4}
\usepackage[intlimits]{amsmath}
\usepackage{amssymb}
\usepackage{psfrag}
\usepackage{graphicx}
\def\be{\begin{equation}}
\def\ee{\end{equation}}
\def\bea{\begin{eqnarray}}
\def\eea{\end{eqnarray}}

\def\C{{\rm\kern.24em
    \vrule width.02em height1.4ex depth-.05ex
    \kern-.26em C}}
\def\N{{\rm I\kern-.18em N}}
\def\R{{\rm I\kern-.21em R}}
\def\Z{{\rm\kern.26em
    \vrule width.02em height0.5ex depth 0ex
    \kern.04em
    \vrule width.02em height1.47ex depth-1ex
    \kern-.34em Z}}
\def\d{{\rm\kern.22em
    \vrule width.02em height1.0ex depth0ex
    \kern-.24em d}}
\def\nn{\nonumber}

\def\C{{\rm\kern.24em
    \vrule width.02em height1.4ex depth-.05ex
    \kern-.26em C}}
\def\N{{\rm I\kern-.18em N}}
\def\O{{\rm\kern.24em
    \vrule width.02em height1.45ex depth-.05ex
    \kern-.26em O}}
\def\P{{\rm I\kern-.25em P}}
\def\R{{\rm I\kern-.21em R}}
\def\Z{{\rm\kern.26em
    \vrule width.02em height0.5ex depth 0ex
    \kern.04em
    \vrule width.02em height1.47ex depth-1ex
    \kern-.34em Z}}

\def\nn{\nonumber}

\newdimen\picraise
\newcommand\picbox[1]
{
  \setbox0=\hbox{\input{#1}}
  \picraise=-0.5\ht0 
  \advance\picraise by 0.5\dp0
  \advance\picraise by 3pt     
  \hbox{\raise\picraise \box0}
}
\newdimen\picraiset
\newcommand\picding[1]
{
  \setbox0=\hbox{\input{#1}}
  \picraiset=-0.5\ht0 
  \advance\picraiset by 0.5\dp0
  \advance\picraiset by -3pt  
  \hbox{\raise\picraiset \box0}
}
\newdimen\picraisehallo
\newcommand\pichallo[2]
{
  \setbox0=\hbox{\input{#1}}
  \picraisehallo=-0.5\ht0 
  \advance\picraisehallo by 0.5\dp0
  \advance\picraisehallo by -#2pt  
  \hbox{\raise\picraisehallo \box0}
}

\begin{document}
\begin{titlepage}
\begin{flushright}
HD-THEP-04-09\\
hep-ph/0403197
\\
\end{flushright}
\vfill
\begin{center}
\boldmath
{\LARGE{\bf The Perturbative Odderon in Quasidiffractive}}
\\[.3cm]
{\LARGE{\bf Photon-Photon Scattering}}
\unboldmath
\end{center}
\vspace{1.2cm}
\begin{center}
{\bf \Large
Stefan Braunewell\,$^a$,  Carlo Ewerz\,$^b$
}
\end{center}
\vspace{.2cm}
\begin{center}
{\sl
Institut f\"ur Theoretische Physik, Universit\"at Heidelberg\\
Philosophenweg 16, D-69120 Heidelberg, Germany}
\end{center}
\vfill
\begin{abstract}
\noindent
We study the perturbative Odderon in the quasidiffractive process 
$\gamma^{(*)} \gamma^{(*)} \to \eta_c \eta_c$. At high energies 
this process is dominated by Odderon exchange and can be viewed 
as the theoretically cleanest test of the perturbative Odderon. 
We calculate the differential and total 
cross section, as well as the dependence on the energy and on 
the photon virtualities taking into account the effects of resummation 
of logarithms of the energy. 
The results are compared with those obtained with a simple 
exchange of three noninteracting gluons. 
We present the expected cross section for this process at a 
future Linear Collider and discuss implications for other 
processes involving the perturbative Odderon. 
\vfill
\end{abstract}
\vspace{5em}
\hrule width 5.cm
\vspace*{.5em}
{\small \noindent 
$^a$ email: S.Braunewell@thphys.uni-heidelberg.de \\
$^b$ email: C.Ewerz@thphys.uni-heidelberg.de 
}
\end{titlepage}

\section{Introduction}
\label{sec:intro}

The Odderon is the partner of the Pomeron carrying negative 
charge parity quantum number. In high energy scattering 
processes it gives the leading contribution to processes in which 
negative $C$-parity is exchanged in the $t$-channel. 

After the concept of the Odderon had been proposed in 
\cite{Lukaszuk:1973nt}, it was for a long time almost exclusively 
discussed in the context of elastic or inclusive processes. These have 
the disadvantage that the Odderon gives only one of many 
contributions to the scattering amplitude and a clean identification 
of the Odderon is rather difficult. The only experimental 
evidence of the Odderon so far has been found as a small difference 
between the differential cross sections of elastic proton-proton 
and antiproton-proton scattering at the CERN ISR \cite{Breakstone:1985pe}. 
Due to the low statistics of the data and the difficulty of 
extracting the Odderon contribution, however, it is not possible 
to interpret this as an unambiguous signal of the Odderon. 
For a more detailed review of the phenomenological and theoretical 
status of the Odderon we refer the reader to \cite{Ewerz:2003xi}. 

Recently an important change of direction in the search for the Odderon 
has taken place. Now the search concentrates on processes 
in which it basically gives the only contribution to the cross section.  
The cross section for such processes is in general smaller than for 
elastic or inclusive processes, but here already the observation 
of the process as such would establish the existence of the Odderon. 
Examples of such exclusive processes are the double-diffractive 
production of vector mesons in proton-(anti)proton scattering 
\cite{Schafer:na} 
or the diffractive production of pseudoscalar or tensor mesons 
in electron-proton scattering 
\cite{Barakhovsky:ra}-\cite{Ma:2003py}. 
In all of these processes Odderon exchange gives the main 
contribution to the cross section at high energies. Other possible 
contributions can only arise due to photon or reggeon exchange, 
but both of these contributions are under good theoretical control. 
Another interesting possibility is to study the interference between 
Pomeron and Odderon exchange. This is possible in the diffractive 
production of final states that can be produced both in a $C=+1$ 
and in a $C=-1$ state like for example a pair of charged pions. 
The interference term between the two corresponding production 
mechanisms can be isolated in suitable asymmetries like for example 
the charge or spin asymmetry. Asymmetries of this kind have been 
studied in \cite{Brodsky:1999mz}-\cite{Ginzburg:2002zd}. 
Also here already the experimental observation of an asymmetry could 
firmly establish the existence of the Odderon. 

The first experimental search for one of these exclusive processes 
was performed for the case of diffractive pion photoproduction in $ep$ 
scattering at HERA in \cite{Adloff:2002dw}. 
This process is the one for which the largest cross section is expected, 
but its theoretical description obviously has to rely on nonperturbative 
techniques. Such a calculation was performed in \cite{Berger:1999ca} 
making use of the stochastic vacuum model 
\cite{Dosch:1987sk,Dosch:ha,Simonov:1987rn} in the framework 
of the functional approach to high energy scattering developed in 
\cite{Nachtmann:1991ua}. 
In \cite{Adloff:2002dw} the experimental results have been compared 
to the expectations based on that calculation, and no signal of Odderon 
exchange has been found. The failure of the theoretical prediction 
for this process is currently not understood. 

In order to avoid the large theoretical uncertainties of nonperturbative 
calculations in diffractive pion production 
one can consider the diffractive production of heavy 
pseudoscalar or tensor mesons. In that case the large mass of the 
meson provides a hard scale, and one can hope that perturbation 
theory is applicable even for real photons. Here in particular 
the production of $\eta_c$ mesons has been considered, 
see \cite{Czyzewski:1996bv,Engel:1997cg,Bartels:2001hw}. 
The expected cross section for that process was in the range 
of several tens of picobarns. 
In a study of elastic $pp$ scattering it has subsequently been found that 
the choice of parameters in the Odderon-proton coupling 
in those calculations was very optimistic \cite{Dosch:2002ai}, 
and a realistic estimate of the cross section should be even 
smaller by at least an order of magnitude. 
Due to the small cross section, the process is not of immediate 
phenomenological interest, but it has turned out to be quite interesting 
from a theoretical perspective. 

That interest is related to the occurrence of large logarithms 
of the energy in the perturbative series. In the simplest possible 
perturbative picture the exchange of an Odderon is described 
by the exchange of three noninteracting gluons in a symmetric 
color state. In higher orders in perturbation theory large 
logarithms of the energy can compensate the smallness of the 
strong coupling constant, $\alpha_s \log s \sim 1$, 
and one needs to resum these logarithms. For the case of the 
Odderon this leads to the generalized leading logarithmic 
approximation (GLLA) which is encoded in the 
Bartels-Kwieci{\'n}ski-Prasza{\l}owicz (BKP) equation. 
Recently two different solutions of this equation have been found 
explicitly in \cite{Janik:1998xj} and in \cite{Bartels:1999yt}. 
The former solution does not couple to the $\gamma \eta_c$ 
impact factor in leading order and is hence not relevant 
for the production of $\eta_c$ mesons. The latter solution, 
the so-called Bartels-Lipatov-Vacca (BLV) solution, on the 
other hand does couple to that impact factor. Its intercept 
exactly equals one, and it hence leads to a cross section which 
is constant with the energy up to logarithmic corrections. 
The BLV solution was recently also found in the dipole picture 
of high energy scattering \cite{Kovchegov:2003dm}. 

Although the intercept of the BLV solution is equal to the 
intercept of the simple three-gluon exchange model for the 
Odderon it has quite different properties. 
So far the phenomenological consequences of using the 
BLV solution in the scattering amplitude have been considered only 
in the diffractive production of $\eta_c$ mesons in 
\cite{Bartels:2001hw,Bartels:2003zu}. Interestingly, in 
\cite{Bartels:2001hw} it was found that for real photons the resulting cross 
section is by about a factor of five larger than the one obtained in 
\cite{Czyzewski:1996bv,Engel:1997cg} by using a simple three-gluon 
exchange for describing the Odderon.
It is a very interesting 
question whether that enhancement is a general property of the 
BLV solution or whether and how strongly it depends on the 
couplings of the Odderon to the proton and to the $\gamma \eta_c$ 
impact factor. It is one of the aims of the present paper 
to address this question. 

All of the processes mentioned above 
involve the coupling of the Odderon to the proton. This coupling 
is known to be rather sensitive to the internal structure of the proton 
\cite{Dosch:2002ai}, and it is therefore possible that due to 
nonperturbative effects this coupling is small. In that case it could 
be quite difficult to find the Odderon in these processes. 
It is therefore interesting to study also processes which do not 
involve the uncertainties of the Odderon-proton coupling. 
From a theoretical point of view the quasidiffractive 
process $\gamma^{(*)} \gamma^{(*)} \to \eta_c \eta_c$ 
is the cleanest possible probe of the Odderon. 
Due to the large mass of the charm quark 
the coupling of the Odderon to the $\gamma \eta_c$ 
impact factor can be calculated perturbatively even for small 
photon virtualities. Again, already the observation of this 
process at high energies would firmly establish the existence 
of the Odderon. More generally, one can study the quasidiffractive 
processes $\gamma^{(*)} \gamma^{(*)} \to M M$ and 
$\gamma^{(*)} \gamma^{(*)} \to M X$ 
with $M$ being a heavy pseudoscalar or tensor meson. 
Also these can occur at high energies only due to Odderon exchange. 
Such processes have first been studied in 
\cite{Ginzburg:gy,Ginzburg:1991hd} and more recently for the case 
of $\eta_c$ meson production in \cite{Motyka:1998kb}. 
In these studies the Odderon has been modeled as a simple exchange of 
three noninteracting gluons. 

In the present paper we study in detail the process 
$\gamma^{(*)} \gamma^{(*)} \to \eta_c \eta_c$ at high energies. 
In particular, we take into account the effects of resumming large 
logarithms of the energy in perturbation theory 
by using the BLV solution of the BKP equation. 
That allows us to perform a detailed study of the 
properties of the BLV Odderon solution in a completely perturbative 
process, that is in a clean theoretical setting which 
does not involve model assumptions about the impact factors. 
The properties of the BLV Odderon can easily be compared to 
those of an Odderon modeled by the exchange of three 
noninteracting gluons. Furthermore, 
by comparing the behavior of the BLV solution in this process 
and in the process $\gamma p \to \eta_c p$ we can draw some 
conclusions about the possible origin of the enhancement obtained 
in the latter process for the BLV solution as compared to simple 
three-gluon exchange. Another important motivation for the 
present study is to estimate the chances of finding the Odderon 
in the process $\gamma^{(*)} \gamma^{(*)} \to \eta_c \eta_c$ 
in $e^+ e^-$ scattering at a future Linear Collider. 

In section \ref{sec:xsect} we provide 
the cross section formulae for the process 
$\gamma^{(*)} \gamma^{(*)} \to \eta_c \eta_c$. In particular 
we discuss the BLV solution and  the $\gamma \eta_c$ impact factor. 
In section \ref{cresults} we study the resulting 
cross section and its dependence on the different 
parameters. After discussing some technical details 
of the calculation, we start with the case of real photons and 
calculate the differential and total cross sections 
in section \ref{scrosssection}. The applicability of the 
saddle point approximation for the BLV Odderon solution 
in this process is considered in \ref{ssaddleresults}. 
We investigate the energy dependence of the cross section 
in section \ref{senergy}. In section \ref{se+e-} we address 
the possibility to observe this process 
at a future Linear Collider. The case of virtual 
photons is studied in section \ref{svirtual}. Finally, we discuss 
our results in the light of results obtained for the BLV solution 
in the process $\gamma p \to \eta_c p$ in section \ref{scompare}. 
Our main results are summarized in section \ref{sec:concl}. 

\section{The scattering amplitude}
\label{sec:xsect}

\subsection{High energy factorization}
\label{shighenergy}

We consider the process $\gamma^{(*)} \gamma^{(*)} \to \eta_c \eta_c$ 
at high energy and relatively small momentum transfer, that is $s \gg |t|$ 
in terms of Mandelstam variables. The photons in the initial state can both 
be real or virtual. The large mass of the charm quark provides a justification 
for treating the process in perturbation theory. 

\begin{figure} 
\begin{center}
\includegraphics[width=7cm]{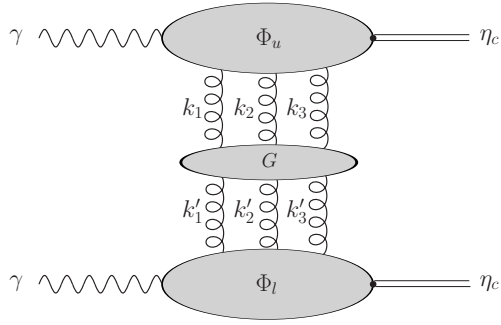}
\caption{Factorized form of the scattering amplitude for 
$\gamma^{(*)} \gamma^{(*)} \to \eta_c \eta_c$ 
\label{amplitudefig}}
\end{center}
\end{figure}

At high energies the process is dominated by Odderon exchange. 
Diagrams involving quark exchange in the $t$-channel are suppressed 
by powers of the energy and can be neglected at the energies which 
we will consider below. Due to high energy factorization the scattering 
amplitude for Odderon exchange can be 
written in the form illustrated in figure \ref{amplitudefig}. 
The amplitude is a convolution of the Odderon Green function 
$\mathbf{G}$ with two impact factors $\Phi$ coupling the 
Odderon to the external particles. Symbolically, 
\be
\label{symbconv}
{\cal A} \sim \langle \Phi_u |\mathbf G | \Phi_l \rangle \,,
\ee
and the convolution includes the integration over the unconstrained 
transverse momenta of the gluons as will be described further below. 
The subscripts $u$ and $l$ of the impact factors stand for the upper 
and lower impact factor, respectively. 
For the Odderon Green function $\mathbf{G}$ 
one can insert either the 
BLV Odderon solution or the propagation of three noninteracting 
gluons as the simplest possible model for the perturbative Odderon. 

In order to make the paper self-contained we collect in the 
following sections the known results 
for the impact factors and for the Odderon Green function, 
and we bring them into a form which can then be used 
to compute the above amplitude and the resulting cross section. 

\subsection{The $\gamma \to \eta_c$ impact factor}

We first consider the impact factor that describes the transition of
a photon (real or virtual) into an $\eta_c$ meson and 
three $t$-channel gluons in a color singlet state as it has been 
calculated in \cite{Czyzewski:1996bv}. There it
was found that the impact factor has only transverse components. 
The resulting expression reads 
\begin{equation}
\label{impfac}
\Phi^i = b \epsilon^i_l \left( \sum_{(123)} 
\frac{(\mathbf{k}_1+\mathbf{k}_2-\mathbf{k}_3)^l}{Q^2+4 m_c^2+
(\mathbf{k}_1+\mathbf{k}_2-\mathbf{k}_3)^2}-\frac{\mathbf{q}^l}
{Q^2+4 m_c^2+\mathbf{q}^2} \right), \qquad i=1,2 \,,
\end{equation}
where the sum runs over all cyclic permutations and the index $i$ corresponds 
to the two possible transverse polarizations of the incident photon. 
Furthermore, $\mathbf{q=k}_1+\mathbf{k}_2+\mathbf{k}_3$ 
is the total momentum transfer, and we will have $t=-\mathbf{q}^2$. 
$Q^2$ is the virtuality of the photon, $m_c=1.4 \, \mbox{GeV}$ 
the mass of the charm quark, $\epsilon_l^i$ the totally antisymmetric tensor 
in two dimensions, and we have 
\begin{equation}\label{phib}
b= \frac{4}{q_c} \frac{d^{abc}}{N_c} \sqrt {\frac{\alpha_s^3}{\alpha} \pi^3 
\Gamma m_{\eta_c}}\,.
\end{equation}
Here $\alpha_s$ and $\alpha$ are the strong and electromagnetic coupling
constants, respectively.  
The charm quark carries the charge $q_c=2/3$, and the $\eta_c$ meson has a
radiative (photon) width of $\Gamma = 7 \, \mbox{keV}$ and a mass of
$m_{\eta_c}=2.98 \, \mbox{GeV}$.

The factor $d^{abc}$ is the totally symmetric structure constant for the color 
SU($N_c$) group. The scattering amplitude contains the contraction 
\begin{equation}
  d^{abc} d^{abc} = \frac{N_c^2-4}{N_c} \,\delta^{aa} = \frac{40}{3} \,,
\end{equation}
where the last equality holds for $N_c=3$. 

\subsection{The BLV solution}
\label{sblv}

The Bartels-Lipatov-Vacca (BLV) Odderon solution 
$\Psi^{(\nu,n)}$ found in \cite{Bartels:1999yt} is constructed 
from the known eigenfunctions $E^{(\nu,n)}$ of the BFKL 
equation \cite{Kuraev:fs,Balitsky:ic}. These eigenfunctions 
are labeled by a discrete quantum number $n \in \Z$, called 
the conformal spin, and a continuous quantum number 
$\nu \in \R$. The functions $E^{(\nu,n)}$ were found 
in impact parameter space in \cite{Lipatov:1985uk}, and 
they can be obtained in transverse momentum space via 
a Fourier transformation. We will use the same symbol $E^{(\nu,n)}$ 
for both representations of the BFKL eigenfunctions. 
In the BLV solution the eigenfunctions of the BKP integral 
operator are constructed as 
\begin{equation}\label{BLVE}
\Psi^{(\nu,n)} (\mathbf k_1, \mathbf k_2, \mathbf k_3) = c(\nu,n) 
\sum_{(123)}  \frac{(\mathbf k_1 +\mathbf k_2)^2}
{\mathbf k_1^2 \mathbf k_2^2} E^{(\nu,n)} 
(\mathbf k_1 + \mathbf k_2, \mathbf k_3) \,, 
\end{equation}
where the sum runs over cyclic permutations, 
and $n$ needs to be an odd integer. 
We use the same normalization convention as in \cite{Bartels:2001hw}, so that the
Odderon states have the same norm as the Pomeron eigenfunctions of which they
are constructed. This leads to 
\begin{equation}
  c(\nu,n)=\sqrt{\frac{N_c \alpha_s}{2 \pi^2 (-3 \chi(\nu,n))}} \,.
\end{equation}
The functions (\ref{BLVE}) are eigenfunctions of the BKP integral operator 
with eigenvalues 
\begin{equation}\label{chi}
  \chi(\nu,n) =\frac{N_c \alpha_s}{\pi} \left [2 \psi(1) - 
\psi\left(\frac{1+|n|}{2}+i \nu\right) -  \psi\left(\frac{1+|n|}{2}-i 
\nu\right) \right ] \,,
\end{equation}
where $\psi$ is the logarithmic derivative of the Euler gamma 
function $\Gamma$. 

The Odderon Green function in spectral representation is constructed as a 
superposition of all states with odd integer numbers $n$ and general 
(real) $\nu$, 
\begin{equation}
\label{BLVG}
 \mathbf G = \sum _\mathrm{odd\ n\ } \int_{-\infty}^{\infty} d\nu \, 
e^{y \chi(\nu,n)} \frac{(2 \pi)^2 (\nu^2+\frac{n^2}{4})} 
{[\nu^2 + \frac{(n-1)^2}{4}]
[\nu^2+\frac{(n+1)^2}{4}]} 
\Psi^{(\nu,n)}
(\mathbf k_1, \mathbf k_2,
\mathbf k_3) {\Psi^{*(\nu,n)}} (\mathbf k_1', \mathbf k_2',
\mathbf k_3') \,.
\end{equation}
Here $y=\log(s/s_0)$ is the rapidity and $s_0$ is a fixed energy scale. 
The scale $s_0$ is undetermined in leading logarithmic approximation, and 
we will discuss possible choices for $s_0$  in section \ref{senergy} below. 
The normalization in (\ref{BLVG}) 
is chosen in such a way that in the limit of vanishing
coupling, $\alpha_s \to 0$, the Green function reduces to the exchange of 
three noninteracting gluons.

In order to calculate the momentum integral for the scattering amplitude,
we need to know the BFKL eigenfunctions in momentum space.
In impact parameter space the eigenfunctions are 
\begin{equation}
  E^{(\nu,n)} (\mathbf r_{10},\mathbf r_{20}) = \left (\frac{r_{12}}
{r_{10}r_{20}} \right )^h \left (\frac{\bar r_{12}}
{\bar r_{10} \bar r_{20}} \right )^{\bar h} \,,
\end{equation}
where $\mathbf r_{ij}= \mathbf r_i-\mathbf r_j$. In the r.h.s.\ we use 
complex coordinates $r_i$ for describing positions in the two-dimensional 
impact parameter space. Further we have the conformal weights 
$h=(1+n)/2+i \nu$ and $\bar h=1-h^*=(1-n)/2+i \nu$. 

The Fourier transform of these eigenfunctions was calculated 
in \cite{Bartels:2001hw}. It was found
that the momentum space functions have the form
\begin{equation}
\label{decanadel}
  E^{(\nu,n)} (\mathbf l_1, \mathbf l_2) =  E_A^{(\nu,n)} (\mathbf l_1, 
\mathbf l_2) +  E_\delta^{(\nu,n)} (\mathbf l_1, \mathbf l_2) \,,
\end{equation}
where $E_A^{(\nu,n)}$ denotes an analytic contribution and
$E_\delta^{(\nu,n)}$ a part containing $\delta$-functions. 
The analytic part reads
\begin{equation}
   E_A^{(\nu,n)}(\mathbf l_1,\mathbf l_2) = C [ 
X(\mathbf l_1, \mathbf l_2) - X(\mathbf l_2, \mathbf l_1) ] \,, 
\end{equation}
where the coefficient $C$ is 
\begin{equation}
 C=\frac{(-i)^n}{(4 \pi)^2} h \bar h(1-h)(1-\bar h)\Gamma(1-h)\Gamma(1-\bar h)
\,. 
\end{equation}
The expression $X$ can
be given in terms of the hypergeometric function 
$F(a_1, a_2; b; z)= {}_2F_1 (a_1, a_2; b; z)$, 
\begin{equation}\label{X}
  X(\mathbf l_1, \mathbf l_2)=\left (\frac{l_1}{2}\right)^{\bar h-2}\left 
(\frac{\bar l_2}{2}\right)
^{h-2} F\left (1-h,2-h;2;-\frac{\bar l_1}{\bar l_2}\right) F\left (1-\bar h,2- \bar h;2;
-\frac{l_2}{l_1}\right) \,.
\end{equation}
The two-dimensional momenta  are denoted as complex numbers on the r.h.s. 

The $\delta$-function part in (\ref{decanadel}) is simpler. 
Denoting the total momentum transfer (in complex notation) by 
$q=l_1 + l_2$, it can be written as 
\begin{equation}
  E_\delta^{(\nu,n)}(\mathbf l_1, \mathbf l_2) = \left [\delta^{(2)}
(\mathbf l_1) + 
(-1)^n \delta^{(2)}(\mathbf l_2) \right] \frac{i^n}{2 \pi} 2^{1-h-\bar h}
\frac{\Gamma(1-\bar h)}{\Gamma(h)} q^{\bar h-1} q^{* h-1} \,. 
\end{equation}

\subsection{Calculation of the scattering amplitude}
\label{sscatteringamplitude}

We want to calculate the scattering amplitudes 
\begin{equation}
\label{scatampl}
 A^{ij} = \frac{s}{3 (2 \pi)^4}\langle \Phi^i_u |\mathbf G | \Phi^j_l \rangle 
\end{equation}
for different transverse polarizations $i,j$ of the incoming photons, 
where we have distributed the constant factors as in \cite{Engel:1997cg}. 
In order to compute these expressions, we have to evaluate integrals 
over the independent transverse momenta, e.g.\ 
\begin{equation}
\langle \Phi^i | \Psi^{(\nu,n)} \rangle = \int d^2\mathbf k_1  
d^2\mathbf k_2 \,\Phi^i (\mathbf k_1,\mathbf k_2;\mathbf q) 
\Psi^{(\nu,n)} (\mathbf k_1,\mathbf k_2;\mathbf q) \,.
\end{equation}
However, this four-dimensional integral reduces to a two-dimensional one 
\cite{Bartels:1999yt}, 
\begin{equation}
\label{reducedintegrals}
\langle \Phi^i | \Psi^{(\nu,n)} \rangle = \frac{b}{c(\nu,n)} \int d^2 {\bf k} \,
\phi^i(\mathbf {k,q-k}) E^{(\nu,n)}\mathbf{(k,q-k)} \equiv \frac{b}
{c(\nu,n)} \langle \phi^i | E^{(\nu,n)} \rangle \,,
\end{equation}
where the reduced impact factor is 
\begin{equation}
\label{reducedimpfac}
\phi^i(\mathbf {k,q-k}) = \frac{\epsilon^i_l (2 \mathbf k - \mathbf 
q)^l} {Q^2 + 4 m_c^2 + (2 \mathbf k - \mathbf q)^2} \,.
\end{equation}

Let us now consider the infinite sum over odd values of $n$ 
in (\ref{BLVG}) which 
needs approximation in order to be evaluated numerically. In the
full Green function (\ref{BLVG}) the exponential factor $e^{y\chi(\nu,n)}$
clearly is of special importance to the integrand, and an expansion
of its argument can help us determine the dominant values of $n$. Expanding
(\ref{chi}) up to second order around $\nu=0$ yields 
\begin{equation}
  \chi(\nu,n) = \frac{N_c \alpha_s}{\pi}\left [ 2 \, \psi(1) -
2 \psi\left(\frac{1+|n|}{2}\right) + \psi''\left(\frac{1+|n|}{2}\right)
\, \nu^2 +
\mathcal O (\nu^4)
\right ] \,.
\end{equation}
For values of $n$ other than $\pm 1$ we therefore get a constant part in the
Taylor expansion of the argument of the exponential which grows with $n$.
We have in fact 
checked numerically the contribution of $n=3$ and find that this
term is already of relative size $\sim 10^{-4}$ compared to the leading term. 
Therefore we can reduce the sum to one over $n=\pm 1$.

Now we can further simplify the 
integral that we have to calculate numerically. The analytic part reduces 
for $n=\pm 1$ to
\begin{equation}\label{reducedanalytic}
  E^{(\nu,n=\pm1)}_A (\mathbf {k, q - k}) = \pm \frac{1}{(4 \pi)^2}\nu(1+\nu^2)
\Gamma^2(1-i \nu)
[X(\mathbf {k, q - k})-X(\mathbf{q - k, k})] \,.
\end{equation}
For $n=1$, (\ref{X}) leads to
\bea
  X^{(n=1)}(\mathbf {k, q - k})  &=&  \left (\frac{k}{2}\right)^{i \nu-2}\left 
(\frac{\bar q - \bar k}{2}\right)^{i \nu-1} 
F\left (-i \nu,1-i \nu;2;-\frac{\bar k}{\bar q - \bar k}\right) 
\nn \\
 &&{}\times \,  F\left (1-i \nu,2-i \nu;2; -\frac{q - k}{k}\right) \,.
\eea
For $n=-1$, we get
\begin{equation}
  X^{(n=-1)}(k, q - k)=X^{(n=1)}(\bar q - \bar k, \bar k) \,,
\end{equation}
and thus obtain
\begin{equation}
  E_A^{(\nu,n=-1)} (k, q - k) =  E_A^{(\nu, n=1)} 
(\bar k,\bar q - \bar k) \,.
\end{equation}

From now on we choose the coordinate axes such 
that the total momentum transfer $\mathbf q$ is in 1-direction, 
so that in complex notation $q$ is purely real. 
As we have an integral over both components of the two-dimensional 
momentum vector $\mathbf k$, we can replace the second component by its 
negative value, which in complex notation corresponds to complex 
conjugation. In the reduced impact factor 
(\ref{reducedimpfac}) we have to switch the sign of the $i=1$ component 
due to the $\epsilon$ symbol. 
The $i=2$ expression remains unchanged because it includes only the 
first component of the momentum in its numerator.
For $i \ne j$ in (\ref{scatampl}) the coherent sum over $n=\pm 1$ thus 
gives two equal but opposite results. 
For parallel photon polarizations, $i=j$, the resulting expression in 
(\ref{scatampl}) is the same for $n=1$ and $n=-1$. Thus, we can work with 
the expression $E_A^{(\nu,n=1)}$, which we will denote as $E_A^{(\nu)}$
from now on.

Let us now turn to the $\delta$-function part for $n=\pm 1$. 
For the $\delta$-function part one also obtains the same 
result for $n=+1$ and $n=-1$, which we will denote by 
$E^{(\nu)}_\delta$, 
\begin{equation}
\label{Enudeltasimp}
E^{(\nu)}_\delta = 
  E^{(\nu,n=\pm 1)}_{\delta}(\mathbf {k, q - k}) = \left [
\delta^{(2)} (\mathbf k)  - \delta^{(2)}(\mathbf{q - k})
 \right] \frac{i}{2 \pi} 4^{- i \nu}
 \frac{\Gamma(1-i \nu)}{\Gamma(1+ i \nu)}\frac{q^{2 i \nu}}
{q} \,. 
\end{equation}
With our specific choice $\mathbf q=(q,0)$ we can easily evaluate
the convolution of the $\delta$-function part with the impact factor. 
We can see from (\ref{reducedimpfac}) that due 
to the $\epsilon$ tensor the $i=1$ component 
of the impact factor vanishes for $\mathbf k = 0$ or $\mathbf {k = q}$, 
that is when the $\delta$-functions in (\ref{Enudeltasimp}) are applied. 
Hence the only non-vanishing contribution to the convolution 
of the $\delta$-function part with the impact factor is the one for 
$i,j=2$, 
\begin{equation}
\langle \phi^{i=2} | E^{(\nu)}_\delta \rangle = \frac{i}{\pi}\frac{1}{4^{i \nu}}
\frac{\Gamma(1-i \nu)}{\Gamma(1+i \nu)} \frac{q^{2 i \nu}}
{Q^2 + 4 m_c^2 + q^2} \,.
\end{equation}

\subsection{Saddle point approximation} 
\label{saddlepoint}

In \cite{Bartels:2001hw} diffractive production of an 
$\eta_c$ meson in $\gamma p$ scattering 
was calculated in the saddle point approximation (SPA). 
For our study this approximation will not be needed, and 
our results given below will not make use of the SPA. 
Nevertheless, it is interesting for us to study the same approximation 
also for our scattering process in order to 
discuss our results in the context of those of \cite{Bartels:2001hw}. 
For this purpose
the $\nu$ integral in (\ref{BLVG}) is approximated by expanding the argument
of the exponential and other $\nu$-dependent factors in the integrand
(in particular the BFKL eigenfunctions) in
Taylor series. In lowest non-vanishing order the Lipatov characteristic 
function (\ref{chi}) is quadratic in $\nu$, 
\begin{equation}
  \chi (\nu, \pm 1) =  -2 \frac{N_c \alpha_s}{\pi} \zeta(3) \nu^2 + \mathcal{O}
(\nu^4) \,.
\end{equation}
The rest of the integrand is also expanded in $\nu$.
The first non-vanishing 
term in the expansion of the analytic part is of first order, but this term 
vanishes in the momentum integral because it is orthogonal to the $\eta_c$
impact factor, see \cite{Bartels:2001hw}. 
The first term that survives this convolution is of second order. 
The $\delta$-function part leads to a non-vanishing contribution already in zeroth
order in $\nu$, 
\begin{equation}
  E^{(\nu)}_{\delta,0} (\mathbf k_1, \mathbf k_2) = \left (\delta^{(2)} 
(\mathbf k_1) - \delta^{(2)} (\mathbf k_2) \right )
\frac{i}{2 \pi}\frac{1}{q} \,.
\end{equation}
Therefore in \cite{Bartels:2001hw} only the $\delta$-contribution is calculated. 
We will study in section \ref{ssaddleresults} below 
how that approximation affects our results. 
As we will show, the SPA turns out to be inadequate for our process. 

\section{Numerical results}
\label{cresults}

\subsection{Details of the calculation}

In the following sections we will present our numerical results 
for the cross section for $\gamma^{(*)} \gamma^{(*)} \to \eta_c \eta_c$ 
and for its dependence on various parameters. In the present 
section we make several general remarks relevant for those results. 

The differential cross section is obtained from the amplitudes $A^{ij}$ 
involving the BLV Odderon solution as 
\begin{equation}
\frac{d\sigma}{dt}=\frac{1}{16 \pi s^2}\frac{1}{4} \sum_{i,j=1}^2 |A^{ij}|^2 
\,. 
\end{equation}
As we have seen in the previous section, the mixed polarization 
amplitudes vanish. Furthermore, we find numerically 
that the $i,j=1$ contribution to the cross section 
is only $\sim 1\%$ of that coming from $i,j=2$. Since the
numerical error of our calculation is also on the percent level 
we neglect this contribution.
We hence have 
\begin{equation}
\frac{d\sigma}{dt}=\frac{1}{64 \pi s^2}|A|^2,
\end{equation}
where $A=A^{22}$.
According to the discussion above we obtain the 
scattering amplitude $A$ as the integral 
\begin{equation}
\label{ampl}
A = 2 \frac{s}{3 (2 \pi)^4} \int_{-\infty}^{\infty}d\nu \,
e^{y \chi(\nu)} \frac{(2 \pi)^2 (\nu^2+\frac{1}{4})} 
{\nu^2(\nu^2+1)} \frac{b^2}{c(\nu)^2} \langle \phi | E^{(\nu)}
\rangle_u \langle \phi | E^{(\nu)} \rangle ^*_l 
\,,
\end{equation}
where we have the reduced impact factor $\phi = \phi^{i=2}$, 
and the convolution of $\phi$ with $ E^{(\nu)}$ is defined in 
(\ref{reducedintegrals}). 
Further, $E^{(\nu)}$ has two parts as in (\ref{decanadel}), 
and we have set $n=+1$ here while multiplying by 2 to 
take into account the contribution of $n=-1$ as explained in the 
previous section. 

Our results below are obtained from a numerical evaluation 
of the integral (\ref{ampl}). We emphasize that we compute 
that integral without further approximations. In particular we 
do not use the saddle point approximation for our main 
results. The outcome of using the SPA is included below 
only in order to discuss the applicability of that approximation. 
In fact we will show that the saddle point approximation is not 
applicable to our process in the phenomenologically relevant 
kinematical region. 

Recall that the convolutions $\langle \phi | E^{(\nu)} \rangle_{u,l}$ 
in the integrand of (\ref{ampl}) involve only two-dimensional 
integrations, see (\ref{reducedintegrals}). The reduction from a 
four-dimensional to a two-dimensional integral in these convolutions 
occurred due to the special structure of the BLV Odderon solution. 
The integral (\ref{ampl}) involves hypergeometric functions which 
are expensive to evaluate in terms of computer time. 
But because of the reduction to two-dimensional integrations in 
$\langle \phi | E^{(\nu)} \rangle_{u,l}$  it is 
still possible to perform the integral (\ref{ampl}) using Mathematica. 

As already pointed out in section \ref{sscatteringamplitude},
we do not take into account contributions to the BLV solution 
with quantum number $n \ne \pm 1$. We have in fact 
calculated numerically the $n=3$ contribution for a variety
of values of the parameters $Q^2$, $s$ and $t$ and find it to 
be negligible. 

One of the most interesting questions which we want to study is 
how the BLV Odderon solution compares to the exchange of 
three noninteracting gluons in our process. The latter exchange 
is the simplest possible perturbative model for the Odderon. 
Technically speaking it amounts to replacing the Odderon Green 
function $\mathbf{G}$ in (\ref{scatampl}) by three free 
gluon propagators, 
\be
\mathbf{G}_{3g} =  \delta^{(2)}(\mathbf{k}'_1- \mathbf{k}_1)
\delta^{(2)}(\mathbf{k}'_2- \mathbf{k}_2) \,
\frac{1}{\mathbf{k}_1^2 \mathbf{k}_2^2 \mathbf{k}_3^2}
\,.
\ee
The cross section for $\gamma^{(*)} \gamma^{(*)} \to \eta_c \eta_c$ 
with a simple three-gluon exchange was calculated in 
\cite{Motyka:1998kb}. We have reproduced the results of that paper 
in order to compare them with our calculations. 

Our calculation is based on the BLV 
Odderon solution which results from the resummation of leading 
logarithms. It should be pointed out that 
there are several uncertainties which are inevitable 
in that approximation scheme. The first of these uncertainties 
concerns the choice of the appropriate value of the strong 
coupling constant. Strictly speaking the scale of $\alpha_s$ is 
undetermined in GLLA. In all our calculations we use 
$\alpha_s=\alpha_s(m_c^2)=0.38$ to allow for an easy comparison with
the results from \cite{Motyka:1998kb} and \cite{Bartels:2001hw} 
where the same value had been chosen. It should be emphasized, 
however, that already a small change in $\alpha_s$ implies 
a considerable change in the cross section. This is due to the simple 
fact that the cross section contains a factor $\alpha_s^6$ already 
from the coupling of the three gluons to the impact factors. 

Another important uncertainty is the choice of the energy scale 
$s_0$ in $y=\log(s/s_0)$. Also this scale is, strictly speaking, 
undetermined in GLLA and has to be chosen as a typical energy 
scale for the process. 
We will discuss several possible choices in detail in section 
\ref{senergy}. It will turn out that choices which appear equally 
natural can lead to quite different results. 

Of course there is a minimal momentum transfer required 
for the transition from a real photon in the initial state to 
a $\eta_c$ meson in the final state. That minimal momentum 
transfer $t_{\mathrm{min}}$ can be estimated to be 
$ t_{\mathrm{min}} \approx - m_{\eta_c}^4/s$.
At the energies which we will consider its numerical value 
will be very small and will not have any quantitative effect visible 
in our figures. 

\subsection{Cross section for the scattering of real photons}
\label{scrosssection}

\begin{figure} 
\begin{center}
\includegraphics[width=14cm]{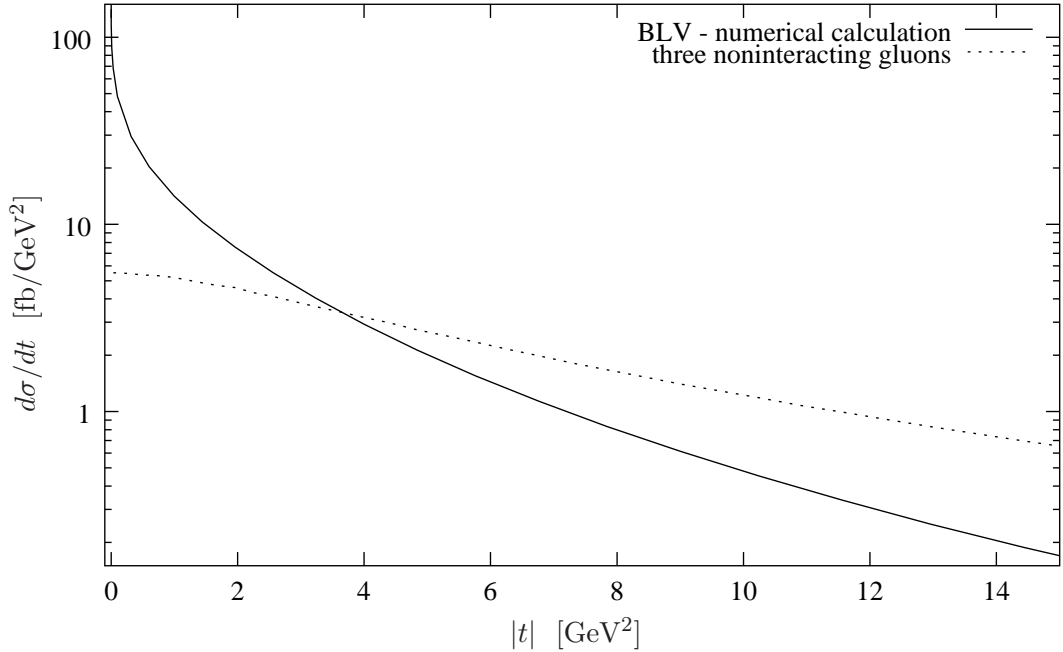}
\caption{Results for the differential cross section $\frac{d\sigma}{dt}$ 
for real photons ($Q^2=0$) and center-of-mass energy $\sqrt{s}=300 
\, \mbox{GeV}$ 
\label{dsigdt}}
\end{center}
\end{figure}

For the calculations in this section we choose a 
center-of-mass energy $\sqrt{s}= 300$ GeV. 
This choice is somewhat arbitrary and our main motivation 
for it is that later on we want to compare the behavior of the BLV 
solution in the process $\gamma \gamma \to \eta_c \eta_c$ 
to that in the process $\gamma p \to \eta_c p$. The latter 
was calculated for the HERA energy in \cite{Bartels:2001hw}, 
and we therefore use the same energy here. 
As the process $\gamma \gamma \to \eta_c \eta_c$ 
is phenomenologically interesting mainly as 
a subprocess of electron-positron scattering for example, 
in an actual collider setup there is a continuous energy range available for the
$\gamma \gamma$ scattering, rather than a fixed center-of-mass energy.
We give an estimate on the size of the cross section for the process
$e^+ e^- \to \eta_c \eta_c$ in section \ref{se+e-}, but in the first part 
we are only concerned with the properties of the differential cross section
of the subprocess.
The squared energy $s$ enters in
the argument of the exponential as $y=\log (s/s_0)$, with the scale factor $s_0$ 
on which we will comment later. For this section we use a scale
$s_0=m_{\eta_c}^2$.
Our numerical results for the differential cross section for
real photons (virtuality $Q^2=Q_{u,l}^2=0$) are shown in figure \ref{dsigdt},
together with the corresponding 
results for the noninteracting three-gluon process from
\cite{Motyka:1998kb}. 

Comparing our numerical results for the BLV solution
to the three noninteracting gluons in the $t$-channel, we find 
a huge enhancement of the differential cross section
at small momentum transfer. The cross section calculated with the 
BLV solution reaches a maximal value of $\approx 120$ 
fb/GeV$^2$ at $t=t_{\mathrm{min}} \approx -10^{-3}\,\mbox{GeV}^2$, 
and then quickly falls to $\approx 88$ fb/GeV$^2$ already at $|t|=0.01$
GeV$^2$. The simple solution exhibits a fundamentally 
different $t$-dependence that can fairly well be described by an exponential
decay. 
Its maximal value at $t=t_{\mathrm{min}}$ is only 
$\approx 5$ fb/GeV$^2$, so there is a
maximal enhancement factor of about $25$ that comes from the interaction of
the three gluons.
As the BLV curve has a much steeper $t$-dependence, the two curves intersect 
at $|t|\approx 3.5 \, \mbox{GeV}^2$.

We have also estimated the total cross sections for the two cases. 
For the calculation involving the BLV solution we get a cross section of about 
${\sigma^\mathrm{BLV}_\mathrm{tot} \approx 59 \, \mbox{fb}}$, whereas
the simple three-gluon process yields 
$\sigma_\mathrm{tot}\approx 43 \, \mathrm{fb}$. 
In both cases,
no cutoff for the integral was needed as the differential cross section falls
off sufficiently quickly with growing $|t|$ to allow for
a reasonable estimate of the contribution from large $|t|$. 
We notice that the minimal momentum transfer $t_{\mathrm{min}}$ is 
sufficiently small at $\sqrt{s}= 300\,\mbox{GeV}$ 
so that it does not affect the calculation of the total cross section. 

In summary, we see that the BLV
solution enhances the total cross section of $\gamma \gamma \to \eta_c \eta_c$ 
by a factor of about $1.5$ compared to the noninteracting three-gluon
calculation of \cite{Motyka:1998kb}. 
The dependence of the differential cross section 
on the momentum transfer changes significantly, and the region of small
momentum transfer becomes more important in the case of the BLV solution. 

\subsection{Comparison with the saddle point approximation}
\label{ssaddleresults}

Next we want to study the reliability of the saddle point 
approximation for our process. This question is primarily of 
theoretical interest. However, it turns out that due to 
the hypergeometric functions in the BLV solution it is numerically 
extremely challenging to calculate processes in which the 
BLV Odderon solution is coupled to a proton. 
For these one has to make use of the SPA, and it is therefore 
interesting to study the reliability of that approximation. 
For this our process is well suited 
since here we can compare the SPA to the exact result. 

\begin{figure} 
\begin{center}
\includegraphics[width=14cm]{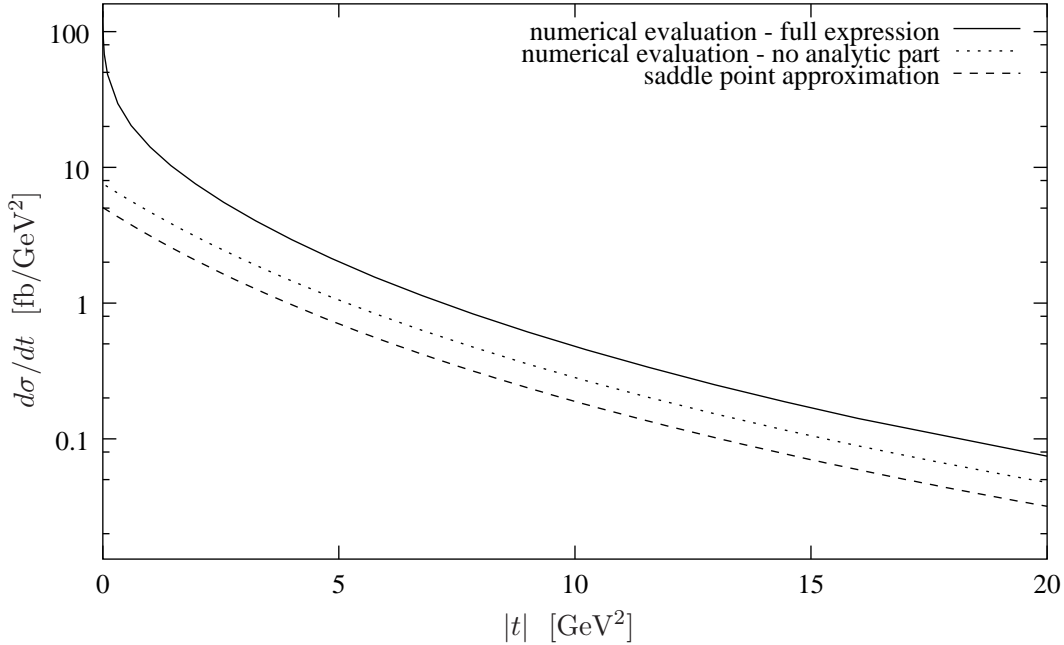}
\caption{Differential cross section: 
comparison of two different approximations of the Odderon wave
function with the exact result 
\label{compsaddle}}
\end{center}
\end{figure}

Figure \ref{compsaddle} shows our exact results together with the
result obtained by using the saddle point approximation as 
described in section \ref{saddlepoint}. Again we have 
chosen $\sqrt{s} = 300\,\mbox{GeV}$. 
We find a maximal
enhancement factor of the exact calculation over the SPA of
order 25 at $t=t_{\mathrm{min}}$, 
but quickly shrinking with increasing $|t|$ to an
enhancement of order 2. For the total cross section the SPA calculation
leads to $\sigma^\mathrm{SPA}_\mathrm{tot} \approx 13 \, \mbox{fb}$, 
hence underestimating the total cross section approximately by a
factor 5. So we find that the SPA in its simplest form 
should not be applied to the scattering process at hand, at least if one is 
interested in small values of $|t|$.

This result appears surprising at first sight, as it was found that 
in calculations involving the BFKL Pomeron 
the saddle point approximation typically overestimates the 
cross section, the deviation from the actual cross section usually being 
of the order of $20 \, \%$, see for example \cite{Bartels:1996ke} 
for the case of the total hadronic cross section in virtual photon 
collisions. We therefore find it instructive to discuss the origin 
of the large deviation and its direction 
in the case of the BLV solution in our process. 

As was mentioned in section \ref{saddlepoint}, 
in the SPA the analytic part $E^{(\nu)}_A$ of the BFKL eigenfunction
is completely neglected.
In figure \ref{compsaddle} we also show how this affects
the $t$-dependence. Going to small $|t|$, the curve 
resulting from the full calculation (including the analytic part)
has a much steeper $t$-dependence than the SPA curve,
leading to the large enhancement. We have
also included in this figure the cross section obtained by neglecting 
the analytic part while calculating the $\nu$ integral numerically 
without the SPA. 
That curve is very similar to the SPA curve in its $t$-dependence, 
but is higher by a factor of about 1.5. 
Thus we see that the crucial difference is not caused by the approximation
of the argument of the exponential, but by the fact that the analytic 
part is neglected.

The omission of the analytic part $E_A^{(\nu)}$ in the SPA 
is due to the fact that the first non-vanishing term in the Taylor 
expansion (in $\nu$) of $E_A^{(\nu)}$ is of second order, whereas the delta
function contribution $E_\delta^{(\nu)}$ 
already has a non-vanishing zeroth order term. But as the factor $y$ in
the exponent is not very large, the analytic piece nevertheless
contributes substantially to the $\nu$ integral. 
Thus a numerical investigation of the $\nu$-dependence of 
the different parts of the solution is needed to obtain a clearer picture 
of the importance of the analytic part. 

\begin{figure} 
\begin{center}
\includegraphics[width=14cm]{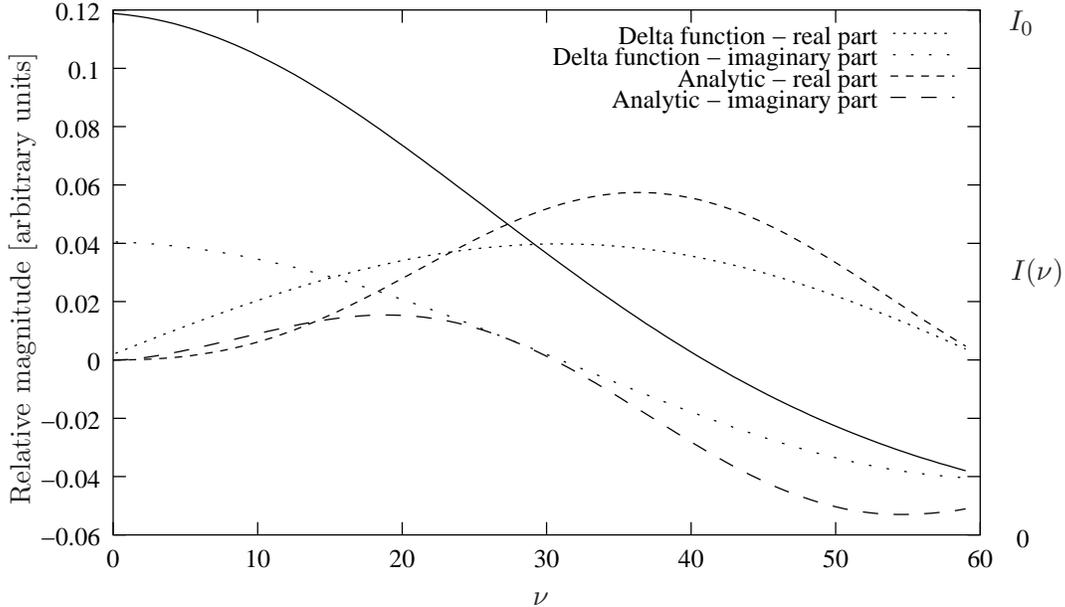}
\caption{Comparison of the different contributions to 
$\langle \phi | E^{(\nu)} \rangle$; the solid line is the Gauss-like function 
$I(\nu)$ representing the momentum-independent factors of the $\nu$-integrand 
(the curve is scaled and shifted, see text) 
\label{fnudep}}
\end{center}
\end{figure}

In figure \ref{fnudep} we
show the real and imaginary parts of both contributions to the expression 
$\langle \phi | E^{(\nu)} \rangle_u$ at $|t|=0.01$ GeV$^2$. 
We have not included the constant factors in this calculation, 
so that the figure shows only the relative size of the 
different terms. We recall that, before being integrated, the expression 
$\langle \phi | E^{(\nu)} \rangle_u$
gets multiplied with the corresponding lower part and the rest of the
$\nu$ integrand in (\ref{ampl}). The latter is given by
\begin{equation}
I(\nu)=e^{y \chi(\nu)} \frac{(2 \pi)^2 (\nu^2+\frac{1}{4})} 
{\nu^2(\nu^2+1)} \frac{b^2}{c(\nu)^2}
\end{equation}
To better understand the significance of a particular contribution to the
overall result, we have included in figure \ref{fnudep} also this 
momentum-independent contribution $I(\nu)$ to the integral (solid curve).
It basically gives a Gauss-like-curve with a maximum $I_0$ at $\nu=0$.
It is scaled and shifted in such a way that the lower horizontal 
axis in the figure is the $I(\nu)=0$ level and the upper one the 
$I(\nu)=I_0$ level (see r.h.s.\ of the figure). 

It now becomes clear from the figure, why the SPA cannot lead to good
results in our calculation. The analytic part (dashed lines) 
gets comparable in size to the $\delta$-function part (dotted) 
already at $\nu \approx 0.2$, where the Gauss-like
factor (solid line) is still at about 70\% of its maximal value. At larger 
values of $\nu$ the analytic part even contributes dominantly to the
amplitude.
Compared to the approximated integrand, where the analytic part is neglected
but the $\nu$ integration is done numerically (no SPA), this gives an
enhancement by a factor of 12 in the differential cross section (which can be
understood when keeping in mind that the expression
$\langle \phi | E^{(\nu)} \rangle$
gets squared when the result of the lower part is multiplied and again squared
when the cross section is calculated).
The analytic part falls off much faster with $|t|$
than the $\delta$-function part, so the SPA improves with increasing momentum
transfer. For values of $|t| > 5 \, \mbox{GeV}^2$ it reproduces the 
$t$-dependence fairly well. Nevertheless, even for larger values of $|t|$ 
the size of the differential cross section is significantly underestimated by the SPA. 
We find that the analytic part $E_A$ becomes negligible at small $t$ 
only for extremely large $y$ above $ \sim 100$. 

\subsection{Energy and scale dependence of the cross section}
\label{senergy}

The intercept of the BLV solution equals unity, so there should be no
power-like energy dependence of the cross section. 
But as the continuous quantum number $\nu$ 
of the BLV solution leads to a cut in the complex angular momentum 
plane (instead of a simple pole), we expect a $\log^{-c(t)} s$ dependence. 
This can be numerically verified
by keeping $t$ fixed and calculating the cross section as a function of $s$. 
Again, the results of this calculation can be compared to the SPA to check 
the significance of the latter. The comparison with the noninteracting three-gluon exchange
process does not give any new insight, as there is no energy dependence in 
that cross section.

The saddle point approximation gives an inverse
logarithmic dependence on the energy, as can be easily seen when keeping 
in mind that $s$ only appears in the Gaussian exponential $\exp(-y c' \nu^2)$ in
the factor $y=\log(s/s_0)$ (with $c'=2 N_c \alpha_s/\pi$).
Similar results are expected for the
numerical calculation if this approximation should be reasonable.
However, as pointed out in the previous section, in the domain of the
momentum transfer that gives the largest contribution to the total cross
section (i.e.\ the small $|t|$ domain) the
applicability of the saddle point approximation is very questionable. 

\begin{figure} 
\begin{center}
\includegraphics[width=14cm]{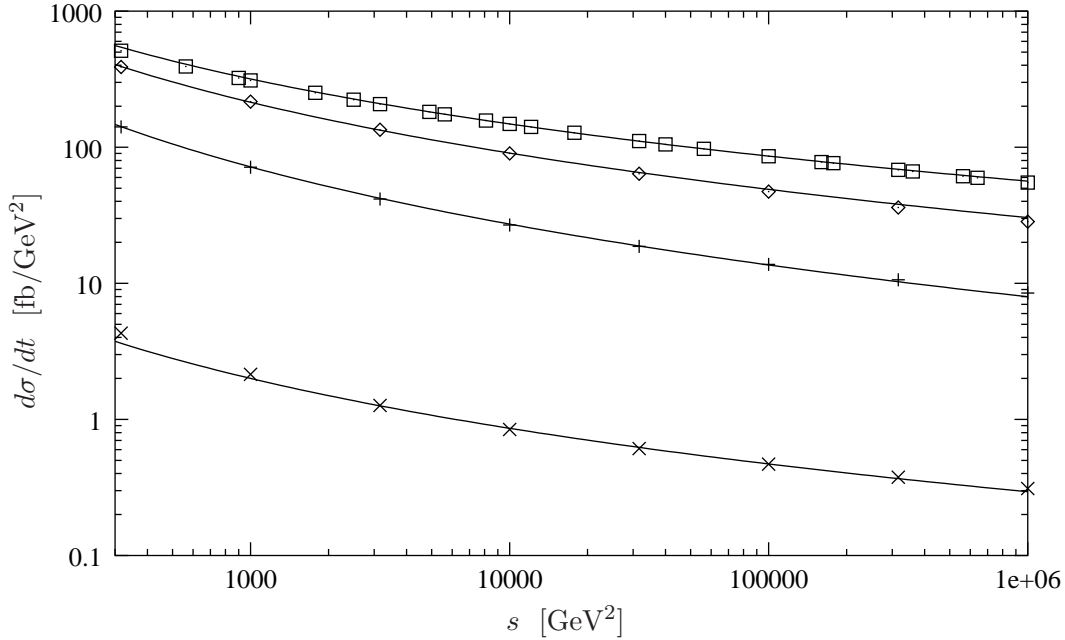}
\caption{$s$-dependence of the differential cross section for real photons 
for $|t|=0.01, 0.1,1,10 \, \mbox{GeV}^2$ (top to bottom), 
the points are numerical results, the lines represent fits (see text) 
\label{fdiffs}}
\end{center}
\end{figure}

Again, we have calculated the differential cross section numerically, 
this time varying $s$. To the results a function of the form 
\begin{equation}\label{fitfunc}
  f(s)=a \log^{-b}(s/s_0) 
\end{equation}
is fitted with fitting parameters $a$ and $b$ that depend only on $t$. 
To see the change of the $s$-dependence with varying $t$, we have 
performed the 
calculation for $|t|=0.01, 0.1, 1$ and $10 \, \mbox{GeV}^2$. The results
for a wide range of squared center-of-mass energies $s$ 
together with the fitted curves are shown in figure \ref{fdiffs}.

One can see that the expected behavior is reproduced quite well by the
numerical results. This should however not be understood as a possibility to 
determine the value of $b$ in (\ref{fitfunc}) as a function of $t$. From 
the fits we get values for $b$ ranging from $1.9$ at $|t|=0.01 \, 
\mbox{GeV}^2$ to $2.4$ at $|t|=1 \, \mbox{GeV}^2$, but in the limit 
of asymptotically large $s$, all these
curves eventually have to approach the saddle point approximation, because
it does become valid at some large value of $y$ when the Gaussian factor gets
so narrow that only the region of small $\nu$ gives a sizeable contribution
to the integrand. Therefore, in the limit $s \to \infty$, we know that 
$b \to 1$. 

In the case of center-of-mass energies accessible at present and planned 
accelerators (up to $y \approx 10$), however, 
the fitted curves give a reasonable
description of the energy dependence. Again, we find the SPA to be 
inappropriate for the parameter range of small $|t|$ and realistic $s$.
In order to keep the figure readable, we did not include the SPA curves,
but again they fail completely in quantitatively reproducing the numerical
results. This can be easily seen from the fact that the exponent of the
logarithm is determined to be around $b \approx 2$, whereas the SPA gives
an exponent $b=1$. It is only at values of $s$ 
well beyond any realistic size ($y \gg 100$) that the saddle 
point approximation becomes acceptable already for small $|t|$. 

We now turn to the question of the dependence of our results
on the choice of the scale $s_0$.
For all of the previous calculations we had chosen the fixed scale 
$s_0=m_{\eta_c}^2$ in $y=\log(s/s_0)$.
The scale $s_0$ of the energy cannot be determined in LLA since 
a change in $s_0$ is formally sub-leading in the expansion of logarithms. 
The numerical results, however, do naturally depend on the specific choice.  
It is therefore interesting to see the influence of different choices of
$s_0$ on the cross section. If $s$, $t$ and $Q$ are fixed, a change of the
scale $s_0$ by a factor $d$ clearly has the same effect as replacing $s$ by
$s/d$ with fixed scale. It is then straightforward to obtain the resulting 
cross sections from the results obtained above. 
As long as the scale $s_0$ remains small compared to $s$, the change does not
qualitatively alter the results. 

Yet, the overall magnitude of the cross section is significantly changed by
a change in the scale factor. For example, in comparison with
$s_0=m_{\eta_c}^2$ the choice $s_0=1 \, \mbox{GeV}^2$, which is the typical
mass scale for hadronic processes, changes the result by an approximately
constant factor of about $2/3$. Compared to this uncertainty, the numerical
and systematical errors in our calculations are definitely negligible.
Strictly speaking, the numerical values of our results should only be taken
as an indication for the order of magnitude. The appropriate choice 
for $s_0$ and hence the absolute results could
only be determined in a next-to-leading order calculation.
The qualitative dependence on the momentum
transfer and on the center-of-mass energy, however, is quite stable under
changes of the scale factor. 

As soon as we consider non-vanishing virtualities of one or both of the 
photons, the choice of the scale $s_0$ becomes even more ambiguous. 
As the
virtuality of the photon provides another momentum scale for the reaction,
it is natural to include it into the scale factor.
The inclusion of the virtuality in $s_0$ will
lead to different results for the dependence of the differential
cross section on the virtuality of the photons, as will be discussed in 
more detail in section \ref{svirtual}.

\subsection{Possible realizations of $\eta_c$ meson photoproduction}
\label{se+e-}

In order to relate our results to phenomenology, we want to give some 
estimates for the cross section in possible future collider setups. So far, 
we have been concerned with
the process $\gamma \gamma \to \eta_c \eta_c$ with real photons 
at the fixed center-of-mass energy $\sqrt{s}=300 \, \mbox{GeV}$.
That scale was chosen having in mind a later comparison to other 
works concerning the process $\gamma p \to \eta_c p$. 

A possible realization of quasidiffractive double  
$\eta_c$ production would be the photon collider option at TESLA. In this
section we use the definitions and numbers given in the TESLA design report 
\cite{Badelek:2001xb}. From an electron-positron
center-of-mass beam energy of $E_\mathrm{beam}=500 \, \mbox{GeV}$ a beam 
of real photons with a 
maximum center-of-mass energy $\sqrt{s}=390 \, \mbox{GeV}$ can be produced.
However, due to the production mechanism of inverse Compton scattering,
the resulting beam is not very narrowly peaked, but has a maximum 
at about $\sqrt{s}=360 \, \mbox{GeV}$ with a width at half 
maximum of about 15 \%. The luminosity for photons with an energy in this 
peak region is estimated as $L_{\gamma \gamma}=1.1 \cdot 10^{34} \mbox{cm}^{-2}
\mbox{s}^{-1}$.
As the dependence on the momentum transfer at this energy does not give
any new insight, we do
not show a figure of the differential cross section
(the curve looks exactly like figure \ref{dsigdt}).
The total cross section for $\sqrt{s}=360 \, \mbox{GeV}$ is 
$\sigma_\mathrm{tot} \approx 55 \, \mbox{fb}$. 
This would lead to a total number of events of the order of $10^5$ in five 
years of continuous running. 

Another possibility of realizing the process $\gamma  \gamma 
\to \eta_c \eta_c$ is directly in electron-positron collisions. 
Here the process occurs as a subprocess in $e^+ e^- \to e^+ e^- \eta_c \eta_c$ 
at high energies. 
The $\gamma \gamma$ subsystem in such a 
collision has a continuous spectrum and we have to integrate
over the energy fractions. In \cite{Motyka:1998kb} this calculation was performed in 
the equivalent photon approximation (see also \cite{Aurenche:1996mz}) 
for the noninteracting three-gluon exchange 
process. There, it was much easier to calculate
the convolution integral, as the simple solution does
not exhibit any energy dependence. Because of the large numerical effort of
calculating total cross sections in our approach, we use the results from
the previous section to estimate the energy dependence of the total cross
section. There we found fairly good fits of the form $\propto \log^b(s/s_0)$
with values of $b$ around 2. Therefore, we approximate the total cross section
by 
\begin{equation}
  \sigma_\mathrm{tot}(s)=\sigma_0 \log^{-2}\left(\frac{s}{s_0}
\right) \,,
\end{equation}
where $\sigma_0=5040 \, \mbox{fb}$ is determined from the value of 
$\sigma_\mathrm{tot}$ at $\sqrt{s}=300 \, \mbox{GeV}$.

The center-of-mass energy $\sqrt{s}$ in the $\gamma \gamma$ subprocess is 
related to the beam energy $E_\mathrm{beam}$ as
\begin{equation}
  s = z_1 z_2 E^2_\mathrm{beam} \,,
\end{equation}
where $z_1$, $z_2$ are the fractions of the electron and positron energies
carried by the two photons. 

In the equivalent photon approximation the energy distribution of the photons
is given by the flux-factor $f_\gamma(z,Q^2_\mathrm{min},Q^2_\mathrm{max})$.
For untagged $e^\pm$ it reads (for details see equations (23)-(25) in 
\cite{Motyka:1998kb}): 
\begin{equation}
  f_\gamma (z) = \frac{\alpha}{2 \pi}\left(\frac{1+(1-z)^2}
{z} \log\frac{(1-z)^2 E^2_\mathrm{beam} \Theta^2_\mathrm{max}}
{m^2_e\, z}\right) \,.
\end{equation}
where we use $\Theta^2_\mathrm{max}=30 \, \mbox{mrad}$. 
The cross section for the process $e^+ e^- \to e^+ e^- \eta_c \eta_c$, which
corresponds to the collision of almost real photons, is given by:
\begin{equation}
  \int_0^1 dz_1 \int_0^1 dz_2 \, \Theta(s-s_\mathrm{min}) \,
\sigma_\mathrm{tot}(s) f_\gamma (z_1) f_\gamma (z_2) \,.
\end{equation}
Here $s_\mathrm{min}$ denotes the minimal squared 
center-of-mass energy for the process. Again, 
$E_\mathrm{beam}= 500 \, \mbox{GeV}$ is used.

If we integrate over the complete domain of possible values for $s$, that is
$4 m_{\eta_c}^2 < s < E^2_\mathrm{beam}$ we get a total cross section for
the electron-positron scattering process of $\approx 55 \, \mbox{fb}$, 
as opposed to $3.5\, \mbox{fb}$ that was
obtained for the simple three-gluon exchange in \cite{Motyka:1998kb}. 
However, with a squared center-of-mass energy $s = s_\mathrm{min} \equiv 
4 m_{\eta_c}^2$ one is clearly not in the high energy limit. In particular,
the requirement $s \gg t$ is not met. Therefore, we have performed the calculation
again for a minimal squared energy of $10 \,s_\mathrm{min}$ and obtain 
for the $e^+ e^-$ total cross section a value of $\sigma_\mathrm{tot} \approx
7 \, \mbox{fb}$. The large difference between these values comes about because
the total photon cross section rises as one goes to small values of $s$. 
Compared to the value at $\sqrt{s}=300 \,\mbox{GeV}$ the total cross section
for $\sqrt{s}=s_\mathrm{min}$ is larger by a factor of 50.

Stating it very cautiously, we estimate the total cross section of
the process $e^+ e^- \to e^+ e^- \eta_c \eta_c$ to be of the order of
$10 \, \mbox{fb}$. The planned luminosity at TESLA is 
$3.4 \cdot 10^{34} \mbox{cm}^{-2} \mbox{s}^{-1}$ \cite{Badelek:2001xb}. 
The resulting order of magnitude of the number of events is similar as in 
the case of the photon collider option. 

A realistic assessment of the feasibility of a measurement of our process 
at a future Linear Collider would clearly require a more detailed study 
of the process including detector cuts and tagging efficiencies. 
Such a study is beyond the scope of the present paper. 

\subsection{Virtual photon scattering}
\label{svirtual}

\begin{figure} 
\begin{center}
\includegraphics[width=14cm]{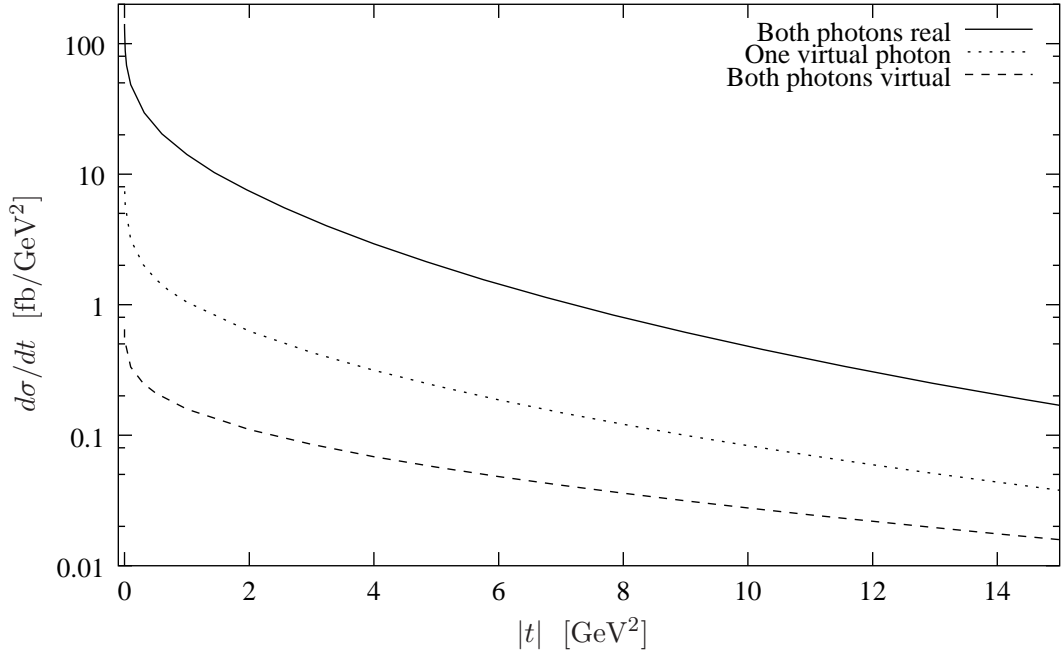}
\caption{Differential cross section for real and virtual 
($Q^2=25 \, \mbox{GeV}^2$) photons for $\sqrt{s} = 300 \, \mbox{GeV}$.
\label{fdifftvarQ}}
\end{center}
\end{figure}

So far we have been concerned with the case of real photon scattering.
In this section we consider the cross section for virtual photons. For
definiteness, we choose a virtuality $Q^2=25 \, \mbox{GeV}^2$, again motivated 
by the choice in \cite{Bartels:2001hw}. 
The comparison of the differential cross section 
for real and virtual photons is presented in figure \ref{fdifftvarQ}. 
We would like to point out that now and in the following we are again 
considering cross sections for $\gamma \gamma$ scattering rather 
than for the corresponding process in $e^+ e^-$ scattering. 
We have included in the figure the cross sections for the
case of two real photons, one real photon and one virtual photon, and two
virtual photons. 

In the latter two processes there is an additional natural scale in the
process that can be included in the scale factor $s_0$. In addition to the
choices discussed in section \ref{senergy} 
this clearly gives another range of reasonable
choices. For the case of two real photons we have used again $s_0=m_{\eta_c}^2$,
for the processes in which at least one virtual photon is involved, we have chosen 
$s_0=m_{\eta_c}^2+Q^2$. That choice appears natural in particular 
in the light of the typical energy scales occurring in deep inelastic scattering. 
In the case of a virtual photon scattering on a real
photon, one could also think of using
some kind of average momentum scale (for example the geometric or the 
arithmetic mean). This would change the 
results only by a factor which is almost independent of $|t|$. 

In figure \ref{fdifftvarQ} it can be seen that the basic dependence on the 
squared momentum transfer is qualitatively similar for all three processes,
but the absolute size of the result is quite different. The virtuality
of a photon leads to a suppression of the differential cross section. 
If both photons are virtual, the cross section is further suppressed.
The relative enhancement of the real photon scattering over the processes 
including virtual photons decreases with growing $|t|$.
This behavior does not come as a surprise if one looks at the reduced impact factor
(\ref{reducedimpfac}) 
\begin{equation}\label{redimpfac}
\phi({\bf k,q-k}) \propto \frac{(2 k^1 - 
q)} {Q^2 + 4 m_c^2 + (2 \mathbf k - \mathbf q)^2} \,.
\end{equation}
The dominant region for the momentum integration
is where the gluon momentum $\mathbf k$ is small.
For small values of $|t|$ the suppression by the virtuality is therefore 
basically given by a factor $\approx 1/Q^2$.
For a value of $|t|$ comparable in size to $Q^2$, the suppression is only 
$\approx 1/2$, and if $|t| \gg Q^2$ the effect of the virtuality 
becomes negligible. 

For the total cross section this gives a strong suppression with respect to
$Q^2$. For one virtual ($Q^2=25\, \mbox{GeV}^2$)
and one real photon, we get a total cross section
$\sigma_\mathrm{tot} \approx 5 \,\mbox{fb}$, for both being virtual 
$\sigma_\mathrm{tot} \approx 1 \,\mbox{fb}$ (for
both real, the cross section was $\sigma_\mathrm{tot} \approx 59\, \mbox{fb}$).
The three-gluon approximation leads to cross sections $\sigma_\mathrm{tot} 
\approx 2\, \mbox{fb}$ for one virtual photon and 
$\sigma_\mathrm{tot} \approx 0.2 \, \mbox{fb}$ if both photons are virtual.
Thus we see that the enhancement of the total cross section 
of the BLV Odderon over the three-gluon approximation gets
amplified when one considers virtual photons.

\begin{figure} 
\begin{center}
\includegraphics[width=14cm]{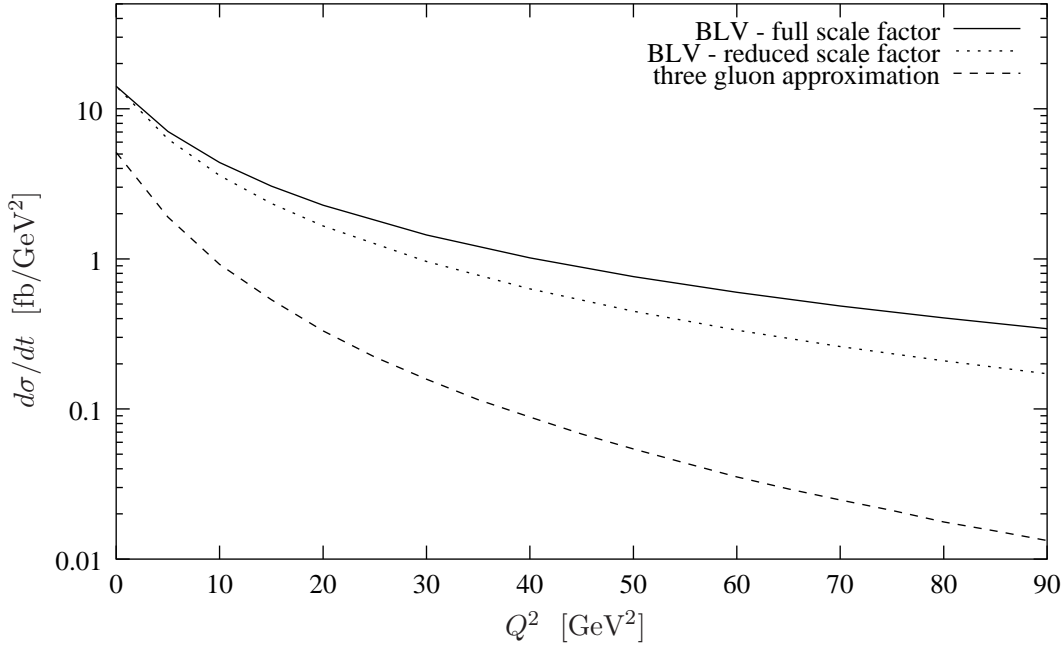}
\caption{$Q^2$-dependence of the differential cross section for 
one virtual and one real photon at 
$|t|=1 \,\mbox{GeV}^2$: comparison of BLV Odderon with 
two different choices for the scale factor $s_0$ 
and noninteracting three-gluon exchange
\label{fQcomp}}
\end{center}
\end{figure}

Next, we want to investigate the dependence of the differential cross section
on the virtuality of one photon (figure \ref{fQcomp}) when the second photon is
real. For this we keep $t$ fixed ($|t|=1 \, \mbox{GeV}^2$)
and vary $Q^2$ in one impact factor. 
As was already mentioned above, 
the results of the numerical calculations depend on the specific
choice of $s_0$ which can now include the virtuality $Q^2$. 
In that case the $Q^2$-dependence of the cross section will 
strongly be affected by the specific choice of $s_0$. 

In figure \ref{fQcomp} we plot the results for three
different calculations: the approximation by three noninteracting
gluons and the numerical BLV Odderon calculation with two choices
for the scale factor. 
These are: the scale 
that was used in \cite{Bartels:2001hw}, 
$s_0=m_{\eta_c}^2+Q^2$, which we will call `full' scale factor because of
the inclusion of the virtuality, and a
`reduced' version without the virtuality of the photon, $s_0=m_{\eta_c}^2$. 
As we have pointed out above, there is no way to determine in LLA 
which scale choice is the correct one. Consequently the terms 
`full' and `reduced' are 
not meant in the sense that the `full' scale factor is `better' in any sense. 
We see that the reduced scale factor leads to a
steeper slope than the full scale factor. Again, we find that the three-gluon 
approximation exhibits a yet steeper slope with respect to
$Q^2$. Nevertheless, all curves have a somewhat similar
dependence on the virtuality.

A remark is in order here concerning the applicability of the BKP Odderon 
in the case of a virtual photon scattering on a real one. If the virtuality 
of the former is large, there is an evolution in transverse momentum 
along the exchanged gluons in the $t$-channel. The BKP equation, 
on the other hand, takes into account only evolution in energy but not 
in transverse momentum. The situation of two largely different 
photon virtualities would hence require the inclusion of DGLAP 
type evolution \cite{Gribov:ri,Altarelli:1977zs,Dokshitzer:sg} and 
is not appropriately described by BKP evolution. This limitation 
of our results for largely different virtualities should be kept in mind 
when interpreting the results of the present section. 

\begin{figure} 
\begin{center}
\includegraphics[width=14cm]{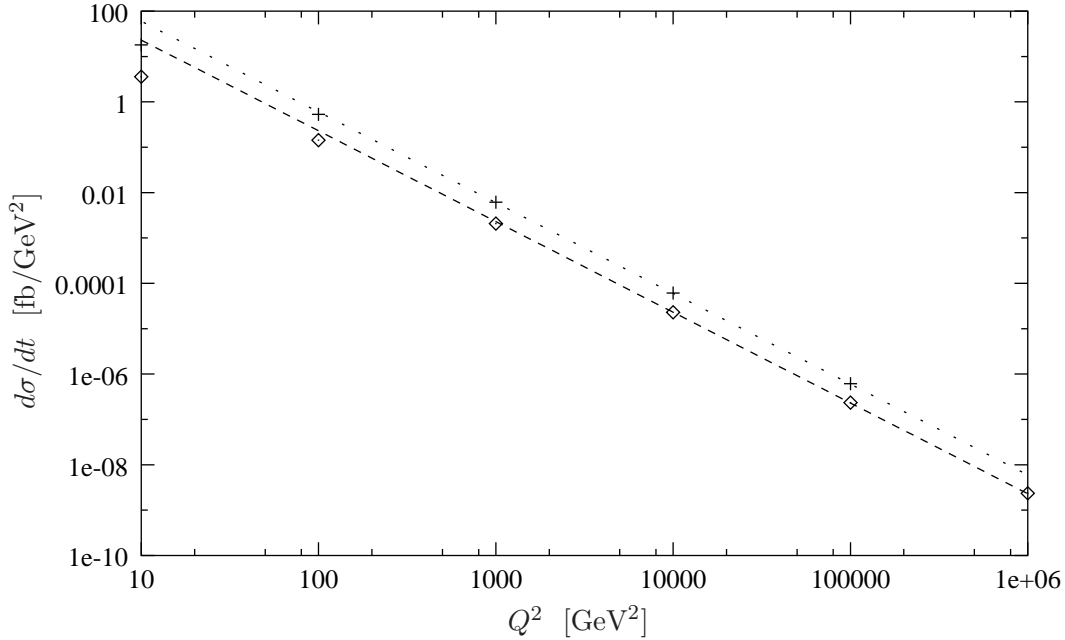}
\caption{$Q^2$-dependence of the differential cross section for 
one virtual and one real photon, $|t|=0.01 \,\mbox{GeV}^2$ 
(upper points) and $|t|=1 \,\mbox{GeV}^2$ (lower points) 
with reduced scale factor $s_0$;
the points are numerical results, the curves are fits (see text) 
\label{fQnoy}}
\end{center}
\end{figure}

\begin{figure} 
\begin{center}
\includegraphics[width=14cm]{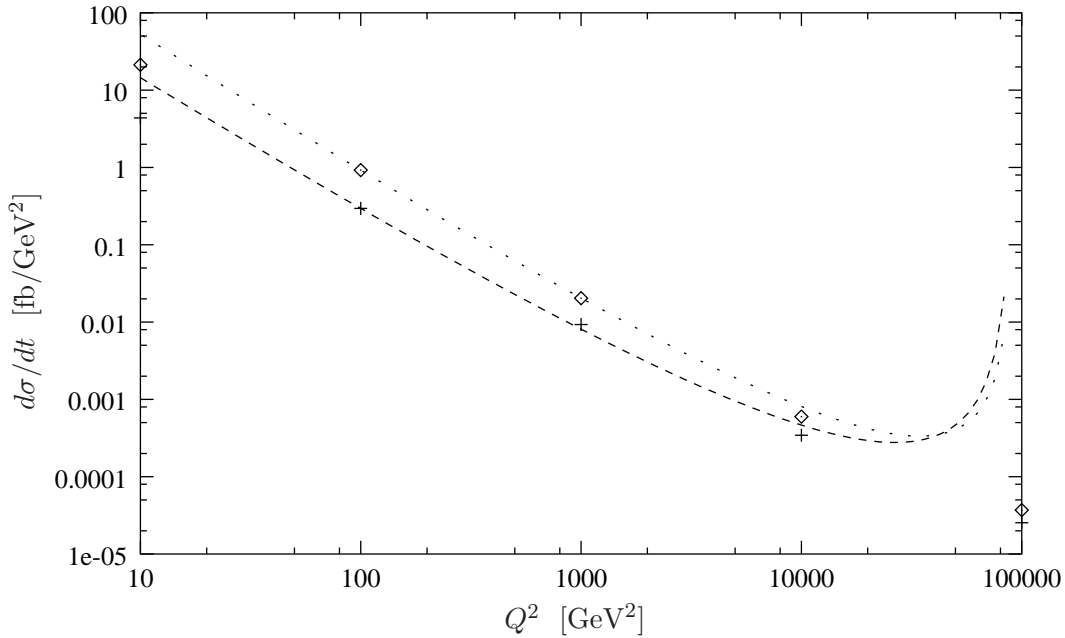}
\caption{$Q^2$-dependence of the differential cross section for 
one virtual and one real photon, $|t|=0.01\, \mbox{GeV}^2$ (upper points) 
and $|t|=1\,\mbox{GeV}^2$ (lower points) with full scale factor $s_0$ 
($Q^2$ included); 
the points are numerical results, the curves are fits (see text) 
\label{fQay}}
\end{center}
\end{figure}

Ignoring that limitation for the moment, we now 
want to investigate the `asymptotic' $Q^2$-dependence. 
That dependence is of theoretical rather than of 
phenomenological interest, but can be useful for gaining 
insight into the BLV Odderon solution. 
Again we consider the case of one virtual and one real photon. 
In figures \ref{fQnoy} (for the reduced scale factor)
and \ref{fQay} (for the full scale factor)
we plot the cross section as a function of $Q^2$ for $|t|=0.01$ and
$1 \, \mbox{GeV}^2$. The points are our numerical results, the
curves are certain fits on which we will comment below. 

If the $Q^2$-dependence is not included in $s_0$, the result is simple. 
In the region $m_c^2 \ll Q^2 \ll s$ we find a ($t$-independent) scaling 
behavior
of the cross section of $\frac{d\sigma}{dt}
\propto Q^{-4}$ (the fitted curves in figure \ref{fQnoy}). 
This can be easily understood. 
The virtuality of the incident photon appears explicitly only in the impact 
factors (\ref{redimpfac}), as the
BLV Green function itself does not exhibit a $Q^2$-dependence. 
The momentum integrals $\langle \phi | E^{(\nu)} \rangle$
receive the dominant contribution from the
region of small transverse momenta. As long as all momentum components are
small compared to $\sqrt{Q^2}$, the error we make by pulling the numerator out of the
integral as a factor $Q^{-2}$ is small. In the cross section this gives
a contribution of $|Q|^{-4}$. We have added to the figures a fit 
with exactly that $Q^2$-dependence to show  
the agreement with the numerical results. 

If we include a $Q^2$-dependence in the scale $s_0$ via the 
`full' scale factor $s_0=m_{\eta_c}^2+Q^2$, the overall
dependence on $Q^2$ changes qualitatively. 
In the domain $m_{\eta_c}^2 \ll Q^2 \ll s$
we basically expect a curve that resembles the $|Q|^{-4}$ result 
from the simple choice $s_0=m^2_{\eta_c}$ multiplied by the 
dependence that we get from section \ref{senergy}, 
$\propto \log^{-b(t)}(s/Q^2)$. In figure \ref{fQay} we show 
the numerical results together with a fit representing that expectation. 
The fit fails when $Q^2$ approaches the boundary 
of the domain in which it is expected to apply but gives 
a fairly good description of the slope within that domain. 
The slope with respect to $\sqrt{Q^2}$ is about 
$3.2$ for $|t|=0.01\,\mbox{GeV}^2$ and about $3.0$ for 
$|t|=1\,\mbox{GeV}^2$, indicating that the slope is not 
entirely independent of $t$. 

A remark is in order here concerning the 
question of the twist of the different solutions to the BKP equation, 
and in particular of the BLV solution. This issue has been 
addressed in \cite{deVega:2002im,Korchemsky:2003rc}, 
see also \cite{Shuvaev:2003gk}. The twist of the Odderon solutions 
is an interesting quantity 
from a theoretical point of view and crucial for a thorough 
understanding of the different solutions. It should be kept in mind, 
however, that at high energies (or small Bjorken-$x$) 
the operator product expansion becomes problematic and 
eventually breaks down, see for example \cite{Mueller:1996hm}. 
Our results illustrate an additional problem of the 
phenomenological aspects of the Odderon twist. 
A potential measurement of the Odderon twist is likely to 
be obscured by the effects of different but theoretically equivalent 
choices of the energy scale $s_0$, at least as long as one works in 
GLLA. 

\subsection{Comparison with photon-proton scattering}
\label{scompare}

In \cite{Bartels:2001hw} the process $\gamma p \to \eta_c p$ 
was studied using the BLV Odderon solution. The coupling of the 
BLV Odderon to the proton is more complicated than the 
$\gamma \eta_c$ impact factor, and here the reduction of the 
four-dimensional integral over two transverse momenta is 
not reduced to a two-dimensional one. 
One therefore has to make use of the saddle point approximation 
for a numerical calculation of the cross section. 
The quality of the saddle point approximation is difficult to 
assess in that process. A comparison with our process can be 
helpful in this respect, since there we can compare the SPA with the 
exact results and are hence able to trace the effects of the 
approximation. In our calculations we have used kinematical 
parameters similar to those of \cite{Bartels:2001hw}, and we can 
therefore hope that our results can give us an idea of what the effects 
of the exact calculation in $\gamma p \to \eta_c p$ can be. 

Before we discuss possible implications of our results for the 
process $\gamma p \to \eta_c p$ we want to point out some general 
differences between the results for that process found in 
\cite{Bartels:2001hw} and the results found here for the 
$\gamma \gamma$ case. 
As was already found in the approximation of three noninteracting 
gluons \cite{Czyzewski:1996bv,Engel:1997cg}
the differential cross section for the process $\gamma p \to \eta_c p$
vanishes as $|t|\to 0$. This is due to the
coupling of the Odderon to the proton impact factor and is not expected
in our scattering process. In addition, the calculation involving the BLV
solution leads to a pronounced dip in the 
cross section at $|t| \approx 0.07 \, \mbox{GeV}^2$.
Also this is due to the coupling of the Odderon to the proton as was 
pointed out in \cite{Bartels:2001hw}, and such a dip is not expected 
in $\gamma \gamma \to \eta_c \eta_c$. 

In \cite{Bartels:2001hw} a sizable difference between the BLV cross section
and the exchange of three noninteracting gluons 
\cite{Czyzewski:1996bv,Engel:1997cg} 
was found in $\gamma^{(*)} p \to \eta_c p$. 
For real photons, the cross section is enhanced by a factor of 5 due to 
resummation, that is for the BLV solution. 
We find a much smaller enhancement in our process. 
In addition, 
when calculating our process in SPA the results are actually 
smaller than for three-gluon exchange. 
Therefore, it appears that in $\gamma p$ scattering 
the enhancement is caused solely by the coupling of the Odderon to
the proton. In the case of virtual photons 
(with $Q^2=25\, \mbox{GeV}^2$) the $\gamma p$ calculation in SPA 
gives an enhancement of the total cross section by one order
of magnitude. This means that there is an additional enhancement factor
of 2 (as compared to the respective three-gluon calculation) 
caused by the virtuality of the photon. We find approximately the same
additional enhancement factor in the case of one virtual photon in 
our process. 

In section \ref{ssaddleresults} we found that for our process the saddle point
approximation significantly underestimates the cross section. 
As the convolution of the Odderon wave function with the $\gamma \eta_c$ 
impact factor in the case of photon-proton scattering was calculated in saddle 
point approximation, too, it is plausible to expect that using the full numerical
calculation also in that process would lead to an additional enhancement that is
approximately the square root of the one found here, i.\,e.\  by about 
a factor 2 for the total cross section. 
We should point out, however, that it is by no means clear that the saddle point 
approximation also underestimates the Odderon-proton coupling. 
Instead, also a completely different behavior of the full 4-dimensional 
integral is conceivable. It would be important to study this problem 
in more detail and to determine at least the direction of the effect of the 
saddle point approximation in that coupling. 

\section{Conclusions and Outlook}
\label{sec:concl}

We have studied the quasidiffractive process 
$\gamma^{(*)} \gamma^{(*)} \to \eta_c \eta_c$ at high energies 
which is mediated by the exchange of an Odderon. 
This process is considered to be the theoretically cleanest 
probe of the Odderon in perturbative QCD since it does not 
involve the uncertainties typically associated with the coupling 
of the Odderon to a proton. 
We have taken into account the effects of resummation of 
large logarithms of the energy by using the BLV Odderon solution. 
This is the only solution of the BKP equation which 
couples in leading order to the $\gamma \eta_c$ impact factor. 

We have investigated in detail the effect of resummation in this 
process by comparing our results to the exchange of three noninteracting 
gluons which is the simplest possible model for a perturbative Odderon. 
For real photons we find that 
resummation strongly enhances the differential cross section at small $|t|$, 
but leads to a faster decrease with increasing $|t|$. The total cross section 
is consequently only slightly enhanced due to resummation. 
The enhancement due to resummation is more pronounced 
when one considers virtual photons. 
We find a logarithmic decrease of the cross section with the energy 
in agreement with the intercept of the BLV solution being exactly one. 
We have discussed in detail the effects of different possible choices 
for the energy scale $s_0$ which is undetermined in leading logarithmic 
approximation. We have investigated this uncertainty and 
find that the cross section is rather sensitive to 
the choice of the energy scale, in particular in the case of virtual photons 
when the scale $s_0$ can naturally involve also the virtuality of the 
photons. 

We have estimated the expected event rates for the process 
$\gamma \gamma \to \eta_c \eta_c$ at a future Linear Collider 
for $e^+e^-$ scattering as well as for a photon collider option. 
In both cases the observation of the Odderon in this process 
appears feasible, but more detailed studies accounting for detector cuts 
and tagging efficiencies will be required to obtain a conclusive assessment 
of this process. 

All previous phenomenological studies of the BLV Odderon solution 
were done for processes involving protons. Due to the complicated 
Odderon-proton coupling the numerical effort for an exact calculation 
of the cross section is prohibitively large in these cases. These processes 
have therefore been studied in the saddle point approximation. 
In our process a considerable simplification occurs due to the special 
structure of the $\gamma \eta_c$ impact factor and an exact numerical 
calculation becomes possible. We have used our exact results to test the 
quality of the saddle point approximation in our process. We find that for 
realistic values of the kinematical parameters the saddle point approximation 
fails and underestimates the actual cross section by about an order of 
magnitude. We have identified the origin of this large deviation and have 
indicated possible implications of our result for processes in which the BLV 
Odderon solution is coupled to a proton. 

Finally, we would like to point out that also other final states can be 
produced via Odderon exchange in quasidiffractive photon-photon 
scattering. An interesting example among them 
is the single-inclusive process $\gamma \gamma \to \eta_c X$ 
which can be treated in a similar way as the process discussed in the 
present paper. In that process, however, the situation is similar to 
the processes in which the BLV Odderon is coupled to a proton. 
The coupling of the BLV Odderon to the $\gamma X$ impact factor 
does not allow one to reduce the four-dimensional integral over 
transverse momenta to a two-dimensional one. An exact numerical 
evaluation of the integral is therefore not feasible and again one 
has to make use of the saddle point approximation. In the light 
of our results, however, it seems likely that also here that approximation 
gives only a relatively poor estimate of the actual cross section. 
Despite this difficulty it would be very interesting to study 
the effects of resummation also in $\gamma \gamma \to \eta_c X$ 
and in related processes involving tensor mesons in more detail 
since they might offer a good chance to observe the Odderon. 

\section*{Acknowledgments}

We would like to thank O.\ Nachtmann and G.\,P.\ Vacca for helpful discussions.


\begin{thebibliography}{99}

\bibitem{Lukaszuk:1973nt}
L.~Lukaszuk and B.~Nicolescu,
Lett.\ Nuovo Cim.\  {\bf 8} (1973) 405.

\bibitem{Breakstone:1985pe}
A.~Breakstone {\it et al.},
Phys.\ Rev.\ Lett.\  {\bf 54} (1985) 2180.

\bibitem{Ewerz:2003xi}
C.~Ewerz,
arXiv:hep-ph/0306137.

\bibitem{Schafer:na}
A.~Sch\"afer, L.~Mankiewicz and O.~Nachtmann,
Phys.\ Lett.\ B {\bf 272} (1991) 419.

\bibitem{Barakhovsky:ra}
V.~V.~Barakhovsky, I.~R.~Zhitnitsky and A.~N.~Shelkovenko,
Phys.\ Lett.\ B {\bf 267} (1991) 532.

\bibitem{Schafer:1992pq}
A.~Sch\"afer, L.~Mankiewicz and O.~Nachtmann,
in Proc.\ of the Workshop "Physics at HERA", Hamburg 1991, vol. 1, p. 243.

\bibitem{Czyzewski:1996bv}
J.~Czyzewski, J.~Kwieci{\'n}ski, L.~Motyka and M.~Sadzikowski,
Phys.\ Lett.\ B {\bf 398} (1997) 400
[Erratum-ibid.\ B {\bf 411} (1997) 402]
[arXiv:hep-ph/9611225].

\bibitem{Engel:1997cg}
R.~Engel, D.~Y.~Ivanov, R.~Kirschner and L.~Szymanowski,
Eur.\ Phys.\ J.\ C {\bf 4} (1998) 93
[arXiv:hep-ph/9707362].

\bibitem{Ryskin:1998kt}
M.~G.~Ryskin,
Eur.\ Phys.\ J.\ C {\bf 2} (1998) 339.

\bibitem{Kilian:1998ew}
W.~Kilian and O.~Nachtmann,
Eur.\ Phys.\ J.\ C {\bf 5} (1998) 317
[arXiv:hep-ph/9712371].

\bibitem{Berger:1999ca}
E.~R.~Berger, A.~Donnachie, H.~G.~Dosch, W.~Kilian, O.~Nachtmann and M.~Rueter,
Eur.\ Phys.\ J.\ C {\bf 9} (1999) 491
[arXiv:hep-ph/9901376].

\bibitem{Berger:2000wt}
E.~R.~Berger, A.~Donnachie, H.~G.~Dosch and O.~Nachtmann,
Eur.\ Phys.\ J.\ C {\bf 14} (2000) 673
[arXiv:hep-ph/0001270].

\bibitem{Bartels:2001hw}
J.~Bartels, M.~A.~Braun, D.~Colferai and G.~P.~Vacca,
Eur.\ Phys.\ J.\ C {\bf 20} (2001) 323
[arXiv:hep-ph/0102221].

\bibitem{Bartels:2003zu}
J.~Bartels, M.~A.~Braun and G.~P.~Vacca,
arXiv:hep-ph/0304160.

\bibitem{Ma:2003py}
J.~P.~Ma,
Nucl.\ Phys.\ A {\bf 727} (2003) 333
[arXiv:hep-ph/0301155].

\bibitem{Brodsky:1999mz}
S.~J.~Brodsky, J.~Rathsman and C.~Merino,
Phys.\ Lett.\ B {\bf 461} (1999) 114
[arXiv:hep-ph/9904280].

\bibitem{Ivanov:2001zc}
I.~P.~Ivanov, N.~N.~Nikolaev and I.~F.~Ginzburg,
in Proc.\ 9th International Workshop on Deep Inelastic Scattering (DIS 2001), 
Bologna, Italy, 2001, 
arXiv:hep-ph/0110181.

\bibitem{Hagler:2002nh}
P.~H\"agler, B.~Pire, L.~Szymanowski and O.~V.~Teryaev,
Phys.\ Lett.\ B {\bf 535} (2002) 117
[Erratum-ibid.\ B {\bf 540} (2002) 324]
[arXiv:hep-ph/0202231].

\bibitem{Hagler:2002nf}
P.~H\"agler, B.~Pire, L.~Szymanowski and O.~V.~Teryaev,
Eur.\ Phys.\ J.\ C {\bf 26} (2002) 261
[arXiv:hep-ph/0207224].

\bibitem{Ginzburg:2002zd}
I.~F.~Ginzburg, I.~P.~Ivanov and N.~N.~Nikolaev,
Eur.\ Phys.\ J.\ directC {\bf 5} (2003) 02
[arXiv:hep-ph/0207345].

\bibitem{Adloff:2002dw}
C.~Adloff {\it et al.}  [H1 Collaboration],
Phys.\ Lett.\ B {\bf 544} (2002) 35
[arXiv:hep-ex/0206073].

\bibitem{Dosch:1987sk}
H.~G.~Dosch,
Phys.\ Lett.\ B {\bf 190} (1987) 177.
 
\bibitem{Dosch:ha}
H.~G.~Dosch and Y.~A.~Simonov,
Phys.\ Lett.\ B {\bf 205} (1988) 339.
 
\bibitem{Simonov:1987rn}
Y.~A.~Simonov,
Nucl.\ Phys.\ B {\bf 307} (1988) 512.

\bibitem{Nachtmann:1991ua}
O.~Nachtmann,
Annals Phys.\  {\bf 209} (1991) 436.

\bibitem{Dosch:2002ai}
H.~G.~Dosch, C.~Ewerz and V.~Schatz,
Eur.\ Phys.\ J.\ C {\bf 24} (2002) 561
[arXiv:hep-ph/0201294].

\bibitem{Bartels:1980pe}
J.~Bartels,
Nucl.\ Phys.\ B {\bf 175} (1980) 365.

\bibitem{Kwiecinski:1980wb}
J.~Kwieci{\'n}ski and M.~Prasza{\l}owicz,
Phys.\ Lett.\ B {\bf 94} (1980) 413.

\bibitem{Janik:1998xj}
R.~A.~Janik and J.~Wosiek,
Phys.\ Rev.\ Lett.\  {\bf 82} (1999) 1092
[arXiv:hep-th/9802100].

\bibitem{Bartels:1999yt}
J.~Bartels, L.~N.~Lipatov and G.~P.~Vacca,
Phys.\ Lett.\ B {\bf 477} (2000) 178
[arXiv:hep-ph/9912423].

\bibitem{Kovchegov:2003dm}
Y.~V.~Kovchegov, L.~Szymanowski and S.~Wallon,
arXiv:hep-ph/0309281.

\bibitem{Ginzburg:gy}
I.~F.~Ginzburg, D.~Y.~Ivanov and V.~G.~Serbo,
Phys.\ Atom.\ Nucl.\  {\bf 56} (1993) 1474
[Yad.\ Fiz.\  {\bf 56N11} (1993) 45].

\bibitem{Ginzburg:1991hd}
I.~F.~Ginzburg and D.~Y.~Ivanov,
Nucl.\ Phys.\ Proc.\ Suppl.\  {\bf 25B} (1992) 224.

\bibitem{Motyka:1998kb}
L.~Motyka and J.~Kwieci{\'n}ski,
Phys.\ Rev.\ D {\bf 58} (1998) 117501
[arXiv:hep-ph/9802278].

\bibitem{Kuraev:fs}
E.~A.~Kuraev, L.~N.~Lipatov and V.~S.~Fadin,
Sov.\ Phys.\ JETP {\bf 45} (1977) 199
[Zh.\ Eksp.\ Teor.\ Fiz.\  {\bf 72} (1977) 377].

\bibitem{Balitsky:ic}
I.~I.~Balitsky and L.~N.~Lipatov,
Sov.\ J.\ Nucl.\ Phys.\  {\bf 28} (1978) 822
[Yad.\ Fiz.\  {\bf 28} (1978) 1597].

\bibitem{Lipatov:1985uk}
L.~N.~Lipatov,
Sov.\ Phys.\ JETP {\bf 63} (1986) 904
[Zh.\ Eksp.\ Teor.\ Fiz.\  {\bf 90} (1986) 1536].

\bibitem{Bartels:1996ke}
J.~Bartels, A.~De Roeck and H.~Lotter,
Phys.\ Lett.\ B {\bf 389} (1996) 742
[arXiv:hep-ph/9608401].

\bibitem{Badelek:2001xb}
B.~Badelek {\it et al.}  [ECFA/DESY Photon Collider Working Group
                  Collaboration],
``TESLA Technical Design Report, Part VI, Chapter 1: Photon collider  at
TESLA,''
arXiv:hep-ex/0108012.

\bibitem{Aurenche:1996mz}
P.~Aurenche {\it et al.},
``Gamma-Gamma Physics at LEP2,''
arXiv:hep-ph/9601317.

\bibitem{Gribov:ri}
V.~N.~Gribov and L.~N.~Lipatov,
Sov.\ J.\ Nucl.\ Phys.\  {\bf 15} (1972) 438 
[Yad.\ Fiz.\  {\bf 15} (1972) 781]. 

\bibitem{Altarelli:1977zs}
G.~Altarelli and G.~Parisi,
Nucl.\ Phys.\ B {\bf 126} (1977) 298.

\bibitem{Dokshitzer:sg}
Y.~L.~Dokshitzer,
Sov.\ Phys.\ JETP {\bf 46} (1977) 641
[Zh.\ Eksp.\ Teor.\ Fiz.\  {\bf 73} (1977) 1216].

\bibitem{deVega:2002im}
H.~J.~de Vega and L.~N.~Lipatov,
Phys.\ Rev.\ D {\bf 66} (2002) 074013
[arXiv:hep-ph/0204245].

\bibitem{Korchemsky:2003rc}
G.~P.~Korchemsky, J.~Kotanski and A.~N.~Manashov,
Phys.\ Lett.\ B {\bf 583} (2004) 121
[arXiv:hep-ph/0306250].

\bibitem{Shuvaev:2003gk}
A.~G.~Shuvaev,
arXiv:hep-ph/0310344.

\bibitem{Mueller:1996hm}
A.~H.~Mueller,
Phys.\ Lett.\ B {\bf 396} (1997) 251
[arXiv:hep-ph/9612251].

\end{thebibliography}
\end{document}